\documentclass[1p,sort&compress,number]{elsarticle}

\usepackage{amssymb}
\usepackage{amsmath}
\usepackage{lscape}
\usepackage{epsf,bezier,eepic}
\usepackage{array}
\usepackage{subfigure}
\usepackage{simplewick}
\usepackage{graphicx}
\newcommand{\bra}[1]{\langle #1 |}

\newcommand{\ket}[1]{| #1 \rangle}
\newcommand{\cket}[2]{|\contraction{}{#1}{}{#2}#1#2 \rangle}
\newcommand{\bck}[3]{\langle #1 | #2 | #3 \rangle}
\newcommand{\clbck}[5]{\langle \contraction{}{#1}{}{#2}#1#2 | #3 | \contraction{}{#4}{}{#5}#4#5 \rangle}
\newcommand{\crbck}[5]{\contraction{\langle}{#1}{#2 | #3 | }{#4} \contraction[2ex]{\langle #1}{#2}{ | #3 | #4}{#5} \langle #1#2 | #3 | #4#5 \rangle}
\newcommand{\bk}[2]{\langle #1 | #2 \rangle}
\newcommand{\be}{\begin{equation}}
\newcommand{\ee}{\end{equation}}
\journal{Annals of Physics}
\begin{document}

\title{Summation of Parquet diagrams as an {\it ab initio} method in nuclear structure calculations}
\author[oslo]{Elise Bergli}
\ead{elise.bergli@fys.uio.no}
\author[oslo]{Morten Hjorth-Jensen}
\ead{morten.hjorth-jensen@fys.uio.no}
\address[oslo]{Department of Physics and Centre of Mathematics for Applications, 
University of Oslo, N-0316 Oslo, Norway}

\date{\today}
\begin{abstract}
In this work we discuss the summation of the Parquet class of diagrams within Green's function theory as a possible framework for {\it ab initio} nuclear structure calculations. The theory is presented and some numerical details are discussed, in particular the approximations employed. We apply the Parquet method to a simple model, and compare our results with those from an  exact solution. The main conlusion is that even at the level of approximation presented here, the results shows good agreement with other comparable {\it ab initio} approaches.
\end{abstract}
\maketitle

\section{Introduction}\label{sec:introduction}

In recent years, considerable progress has been made in the field of nuclear structure calculations. Due to the increasing computational power available with better and faster super-computers, the regime where {\it ab initio}\footnote{With {\it ab initio} we mean here an approach where the many-body Schr\"odinger equation is solved with no approximation given a non-relativistic Hamiltonian. Many-body methods with mathematically controllable truncations fall also under this category as well.} calculations are possible has increased considerably. For lighter nuclei, several approaches have been successful, including the coupled-cluster method \cite{coester,coesterkummel,Bartlett2007,bishop97,Dean2004a,Hagen2007,Hagen2008}, variational approaches \cite{Varga1995,Viviani2005}, large-scale diagonalization techniques (including the no-core shell model) \cite{Caurier2005}, the effective interactions hyperspherical method \cite{Gazit2006} and the Green's function Monte Carlo method \cite{Pudliner1997,Koonin1997,Ceperley1995}. For nuclei with $A\leq{4}$, the Faddeev/Faddeev-Yakubovsky equations \cite{Nogga2000} provide almost exact numerical solutions, and such calculation are used as a benchmark for testing other approaches. The Green's function Monte Carlo method provides an almost exact reproduction of the Faddeev results \cite{Arriaga1995}, but the scaling of the computational effort to the number of particles involved makes this approach impractical for medium-sized nuclei, the current limit being at about $A\lesssim{12}$. The coupled-cluster (CC) method also shows excellent agreement with the Faddeev/Faddeev-Yakubovsky results for $^3$H and $^4$He \cite{Hagen2007}, and this method scales much more favorably, currently being applied to 
nuclei with $A\lesssim{48}$ \cite{Hagen2008}. 

Another method with strong potential for {\it ab initio} calculations of medium-mass nuclei is the self-consistent Green's functions approach \cite{Barbieri2006,Barbieri2009}. This method has a number of interesting features. It is possible to obtain better accuracy with a smaller numerical effort when compared to large-scale diagonalization approaches, and the self-consistency requirement provides a path to ensure the conservation of basic macroscopic quantities. The main advantage, however, is the close connection with experimentally available data sets through the elementary basic blocks of the theory, namely the many-body propagators. These contain information on several excitation processes. The single-particle spectral function (and spectroscopic factor) can be experimentally extracted, for example in $(e,e^{\prime})$ knock-out reactions \cite{Lapikas1993,Leuschner1994}, see also Refs.~\cite{Subedi2008,Rohe2004,Barbieri2004a}. Baym and Kadanoff \cite{Baym1961,Baym1962} showed that the self-consistency requirement ensures the conservation of basic quantities like number of particles, energy, momentum and angular momentum. Recently, converged results for the ground state energy of ${}^4$He~\cite{Barbieri2009b} within the self-consistent Green's function framework employed by Barbieri {\it et.al.}\cite{Barbieri2004,Barbieri2006,Barbieri2009} have been presented. These results show that the required level of precision to meet the current standard is possible within implementations of Green's function based methods.

In this paper we explore the performance of the Parquet method of re-summation of diagrams within the Green's functions theory to calculate the ground state energy in a simplified model of the nucleus. We compare our results with an exact diagonalization. The aim of the paper is to ascertain the possibilty of establishing the Parquet summation method as an alternative {\it ab initio} method for nuclear structure calculations.

The Parquet method of summing diagrams has been known for more than 50 years, having first been developed by Diatlov, Sudhakov and Ter-Martirosian \cite{Diatlov1957} as an aid to describe meson-meson scattering in particle physics. These equations have since been used somewhat, most notably in one- and two-dimensional electron gas calculations \cite{Zheleznyak1997,Janis2006}. They have also been used for some critical-phenomena calculation \cite{Bickers1992,Yeo1996}. The most exhaustive theoretical investigations were carried out by Jackson, Land\'{e} and Smith \cite{Jackson1982,Jackson1994,Jackson1985,Lande1992}. More recently, Yasuda has used the Parquet diagram method to construct approximations to the reduced density matrix of general quantum systems \cite{Yasuda1999}. 

The effective interaction generated by this method includes a large class of diagrams, and requires no initial assumptions on the underlying interaction with respect to range and strength, as opposed to ladder type or ring type (standard random-phase approximation) interactions. The generated interaction is symmetric, that is, the particle-particle part of the interaction and the particle-hole part is treated on an equal footing, thus ensuring that all diagrams critical to a reasonable description of the many-body system are included. It is therefore applicable to systems where easier approaches fail, for example systems where it seems that both particle-particle type (calculated by ladder/$G$-matrix approximations \cite{Ellis1977,mhj1995}) and particle-hole type (usually handled by random-phase approximation (RPA) methods) diagrams are equally important. Systems undergoing a phase transition consist of one such example, and there is clear evidence that both these parts of the interactions play a crucial role in nuclear systems \cite{Ellis1977,mhj1995}. By making a self-consistent calculation, the included diagram classes are all automatically summed to all orders. Only linked diagrams are included in the sum, which ensures that the method is size extensive, meaning that the total energy scales correctly with the number of particles \cite{Bartlett2007}. 

As mentioned above, the aim of this work is thus to demonstrate how one can sum the Parquet class of diagrams in a simple model and compare these results with other {\em ab initio} methods such as diagonalization methods, with a critical eye on the pros and cons of the method. As such, this article serves mainly as a proof-of-principle  of the Parquet method for nuclear-physics like 
systems where pairing correlations and core-polarization play important roles. 

This work is organized as follows: in Section \ref{sec:manybody} we review some basic features of many-body theory and Green's function theory, before presenting our description of the Parquet summation method in Section \ref{sec:parquet}. In Section \ref{sec:itersolution} details on the numerical procedure and approximations are discussed. We present comparisons between our calculations and the exact diagonalization solution for a simple model in Section \ref{sec:simple}, and then we summarize our findings in Section \ref{sec:conclusions}.

\section{Green's  Functions and the Interaction operator} \label{sec:manybody}

Parquet theory is formulated in the language of Green's function theory (or propagator formalism). This formalism is a standard framework within many-body quantum theory, and fuller accounts are found in most textbooks on the subject, for example in Refs.~\cite{FetterWalecka1971,Mattuck1976,BlaizotRipka1986} or \cite{Abrikosov1965}. A comparatively recent presentation is given by the text of Dickhoff and Van Neck from 2004 \cite{Dickhoff2005}. Here we shall be content with a short discussion of the one- and two-body propagators (Section \ref{subsec:onetwobodyprop}), the interaction operator (Section \ref{subsec:Gamma4pt}), the self energy (Section \ref{subsec:selfenergy}), and finally, the matrix inversion method of finding the one-body propagator (Section \ref{subsec:sp_matinv}). We refer the reader to the above-mentioned references for an introduction to the basic concepts and for further details.

\subsection{Propagators} \label{subsec:onetwobodyprop}

The one-particle Green's function is defined as:
\begin{equation} \label{eq:1p_prop_tau}
\begin{split}
g_{\alpha{\beta}}(\tau)=g_{\alpha{\beta}}(t-t') & =-i\langle{\Psi_{0}^{N}|\mathcal{T}\{c_{\alpha}(t)c_{\beta}^{\dagger}(t')\}|\Psi_{0}^{N}\rangle} \\
&= \begin{cases}
  -i\langle{\Psi_{0}^{N}|c_{\alpha}(t)c_{\beta}^{\dagger}(t')|\Psi_{0}^{N}\rangle} & t>t' \\
  i\langle{\Psi_{0}^{N}|c_{\beta}^{\dagger}(t')c_{\alpha}(t)|\Psi_{0}^{N}\rangle} & t\leq{t}'.
  \end{cases}
\end{split}
\end{equation}
Here $c_{\alpha}(t)$ and $c_{\beta}^{\dagger}(t')$ are the annihilation and creation operators in the Heisenberg representation, $\ket{\Psi_{0}^{N}}$ is the N-particle ground state and $\mathcal{T}$ is the time ordering operator. If we add a particle in state $\beta$ at a given time $t'$, the one-body propagator for $t>{t}'$ gives the probability that we find the system still in its ground state if we remove a particle in state $\alpha$ at time $t$. Similarly, for $t{\le}t'$ the one-body propagator gives the probability for recovering the ground state when a hole is created (a particle removed) at a time $t$ and then annihilated at ${t}'$. Fourier transforming to obtain the so-called Lehmann representation, we see that the denominator is zero at energies corresponding to the excitation energies of the $(N+1)$ and the $(N-1)$ states with respect to the ground state $\ket{\Psi_{0}^{N}}$ \cite{Dickhoff2004}:
\begin{equation} \label{eq:1p_prop_omega}
\begin{split}
g_{\alpha{\beta}}(\omega) & = \frac{1}{2\pi{}}\int_{-\infty}^{\infty}d\tau{e^{i\omega{\tau}}}g_{\alpha{\beta}}(\tau) \\
& = \sum_{n}\frac{\bck{\Psi_{0}^{N}}{c_{\alpha}}{\Psi_{n}^{N+1}}\bck{\Psi_{n}^{N+1}}{c_{\beta}^{\dagger}}{\Psi_{0}^{N}}}{\omega-(E_n^{N+1}-E_0^N)+i\eta} + \sum_{k}\frac{\bck{\Psi_{0}^{N}}{c_{\beta}^{\dagger}}{\Psi_{k}^{N-1}}\bck{\Psi_{k}^{N-1}}{c_{\alpha}}{\Psi_{0}^{N}}}{\omega-(E_0^{N}-E_k^{N-1})-i\eta} \\
 & \equiv \sum_{n}\frac{z_{\alpha{\beta}}^{n+}}{\omega-\epsilon_n^{+}+i\eta} + \sum_{k}\frac{z_{\alpha{\beta}}^{k-}}{\omega-\epsilon_k^{-}-i\eta},
\end{split}
\end{equation}
where the last equation introduces the notation $z_{\alpha{\beta}}^{n+}$ as abbreviation for \\
$\bck{\Psi_0^{N}}{c_{\alpha}}{\Psi_n^{N+1}}\bck{\Psi_n^{N+1}}{c_{\beta}^{\dagger}}{\Psi_0^{N}}$ and so on. The energies $\epsilon_n^{+}$ and $\epsilon_k^{-}$ are the energy differences $E_n^{N+1}-E_0^N$ and $E_0^{N}-E_k^{N-1}$ respectively.

The unperturbed (non-interacting, or free) one-particle propagator is given by:
\begin{equation}\label{eq:1p_prop_free}
g_{\alpha{\beta}}^{0}(\omega)=\delta_{\alpha,{\beta}}\Bigl{(}\frac{\theta(\alpha-F)}{\omega-e_{\alpha}^{0}+i\eta}+\frac{\theta(F-\alpha)}{\omega-e_{\alpha}^{0}-i\eta}\Bigr{)},
\end{equation}
where $F$ is the highest occupied state (at the Fermi level) in the system and $e_{\alpha}^{0}$ is the unperturbed energy of the state $\ket{\alpha}$. In this case the energy differences between the energy of the state with $N$ particles and the states with $N\pm{1}$ particles is just the energy of the single-particle state added or removed. 


To study the effects of interactions between the single-particle states, the following representation of the diagonal elements of the single-particle propagator is useful
\be
g_{\alpha{\alpha}}(\omega) = \int_{-\infty}^{\infty}\frac{d\omega'}{2\pi}\frac{S({\alpha},\omega')}{\omega'-\omega}.
\ee
Here $S({\alpha},\omega)$ is the spectral function, given by
\be
S(\alpha,\omega) = -i\lim_{\eta{\to}0_{+}}[g_{\alpha{\alpha}}(\omega+i\eta)-g_{\alpha{\alpha}}(\omega-i\eta)]. 
\ee
The hole part of this is valid for energies $\omega$ less than the lower Fermi energy $\epsilon_{F}^{-}=E_{0}^{N}-E_{0}^{N-1}$, and is given by  
\begin{equation}\label{eq:Sh}
\begin{split}
S_{h}(\alpha,\omega) & = \frac{1}{\pi}\text{Im}g_{\alpha{\alpha}}(\omega) \\
 & = \sum_{n} |\bck{\Psi_{n}^{N-1}}{c_{\alpha}}{\Psi_{0}^{N}}|^{2}\delta(\omega-(E_{0}^{N}-E_{n}^{N-1})).
\end{split}
\end{equation}
This quantity gives the probability at a given energy $\omega$ of removing a particle (creating a hole) with quantum numbers $\alpha$ while leaving the remaining $N-1$ particle system at an energy $E_{n}^{N-1}=E_{0}^{N}-\omega$. Similarly, the particle part $S_{p}(\alpha,\omega)$ is valid for energies $\omega>\epsilon_{F}^{+}=E_{0}^{N+1}-E_{0}^{N}$. It is given by 
\begin{equation}
\begin{split}
S_{p}(\alpha,\omega) & = -\frac{1}{\pi}\text{Im}g_{\alpha{\alpha}}(\omega) \\
 & = \sum_{m} |\bck{\Psi_{m}^{N+1}}{c_{\alpha}^{\dagger}}{\Psi_{0}^{N}}|^{2}\delta(\omega-(E_{m}^{N+1}-E_{0}^{N})),
\end{split}
\end{equation}
and is the probability for adding a particle with quantum numbers $\alpha$ to an $N$-particle system with energy $\omega$, resulting in an $N+1$-system with energy $E_{n}^{N+1}=E_{0}^{N}+\omega$. 

For a given single-particle state, we can define the occupation number $n(\alpha)$ and the depletion number $d(\alpha)$ as 

\begin{equation}
n(\alpha) = \bck{\Psi_{0}^{N}}{c_{\alpha}^{\dagger}c_{\alpha}}{\Psi_{0}^{N}} = \int_{-\infty}^{\epsilon_{F}^{-}}d\omega \quad S_{h}(\alpha,\omega),
\end{equation}
and
\begin{equation}
d(\alpha) = \bck{\Psi_{0}^{N}}{c_{\alpha}c_{\alpha}^{\dagger}}{\Psi_{0}^{N}} = \int^{\infty}_{\epsilon_{F}^{+}}d\omega \quad S_{p}(\alpha,\omega),
\end{equation}
respectively. It can be shown that $n(\alpha) + d(\alpha) = 1$, see for example Ref.~\cite{Dickhoff2005}.

In a non-interacting system, choosing the set $\ket{\alpha}$ determined by the single-particle Hamiltonian $H_{0}$ as the basis gives the hole and particle spectral functions a particularly simple form, being delta functions with height 1 at the energies corresponding to the eigenvalues of the single-particle Hamiltonian. 

In interacting systems, the spectral functions become smeared out. In principle, the number of poles in the propagator is infinite, giving a continuous distribution of probabilities for the energies of the $N\pm{1}$ particle systems. As long as the independent-particle picture remains relatively correct (that is, if the interactions between the particles are weak), the spectral functions will have sharp peaks at clearly defined energies, which we then identify as single-particle states.

The hole spectral function is relatively easy to compare to experimental data extracted from knock-out ($e,e^{\prime}p$) reactions in nuclei \cite{Lapikas1993,Leuschner1994}. In these experiments, an incident fast electron transfers a large amount of energy to a single proton inside the nucleus, sufficient to eject the proton, and the momentum profiles of this proton and the scattered electron are then measured. The most commonly extracted quantity is the so-called spectroscopic factor, defined as 
\begin{equation}
  S_{\alpha} = \int{d{\bf p}}|\bck{\Psi_n^{N-1}}{a_{\bf p}}{\Psi_0^{N}}|^2,
\end{equation}
where $a_{\bf p}$ is a momentum state annihilation operator. In an independent-particle system the spectroscopic factor is either 0 (unoccupied state) or 1 (occupied state). When the spectroscopic factor is less than 1, it can be thought of as measuring the amount of correlation present in the $N$-particle system, being the difference between the independent-particle spectroscopic factor of 1 and the measured value. 
A word of caution is however necessary here. 
Experimentally, spectroscopic factors are defined as the ratio
of the observed reaction rate with respect to the same rate calculated
assuming a full occupation of the relevant single-particle states.
They are therefore often interpreted as a measure of the occupancy
of a specific single-particle state. 
However, from a strict theoretical point of view spectroscopic factors 
are not occupation numbers but a measure of what fraction of
the full wave function can be factorized into a correlated state (often
chosen to be a given closed-shell core)  and an independent single-particle
or single-hole state. Large deviations from the values predicted by
an independent-particle model, point to a strongly correlated system.

In our formalism, the spectroscopic factor is given by the height of the spectral function at the energy of the $\ket{\Psi_n^{N-1}}$ state (this follows from the orthogonality of the basis set $\ket{\alpha}$). 


From applying the equation of motion for a Heisenberg operator, $dc_{\alpha}(t)/dt=-i[c_{\alpha}(t),\hat{H}]$ to Eq.~(\ref{eq:1p_prop_tau}), one obtains the first step in the Martin-Schwinger hierarchy \cite{Martin1959}, relating the N+1-particle propagator to the N-particle propagator. Thus, relating the two-particle propagator to the one-particle propagator \cite{Abrikosov1965}:
\begin{equation} \label{eq:eq_mot_1prop}
\begin{split}
i{}\frac{\partial}{\partial{t}}g_{\alpha{\beta}}(t-t')& = \quad \frac{\partial}{\partial{t}}\bck{\Psi^{N}_0}{\mathcal{T}[a_{\alpha_H}(t)a^{\dagger}_{\beta_H}(t^{\prime})]}{\Psi^{N}_0} \\
 & =\quad\delta(t-t')\delta_{\alpha,{\beta}}+\epsilon_{\alpha}g_{\alpha{\beta}}(t-t') \\
 &+ \frac{-i}{2}\sum_{\eta{\gamma}\sigma}\bck{{\alpha}\eta}{V}{\gamma{\sigma}}\langle{\Psi_{0}^{N}|\mathcal{T}\{c_{\eta}^{\dagger}(t)c_{\sigma}(t)c_{\gamma}(t)c_{\beta}^{\dagger}(t')\}|\Psi_{0}^{N}\rangle}.
\end{split}
\end{equation}
This generates a term containing the 4-point Green's function, defined by
\begin{equation} \label{eq:4pointgreen}
\begin{split}  
K_{\alpha{\beta},\gamma{\delta}}(t_{\alpha},t_{\beta};t_{\gamma},t_{\delta})&=-i{}\langle{\Psi_{0}^{N}|\mathcal{T}\{c_{\beta}(t_{\beta})c_{\alpha}(t_{\alpha})c_{\gamma}^{\dagger}(t_{\gamma})c_{\delta}^{\dagger}(t_{\delta})\}|\Psi_{0}^{N}\rangle} \\
&\equiv \bck{\alpha{\beta}}{K(t_{\alpha},t_{\beta};t_{\gamma},t_{\delta})}{\gamma{\delta}}.
\end{split}
\end{equation}
Since the  4-point Green's function is antisymmetric under exchange of indices, it is possible to define matrix elements of $K$ between antisymmetric two-particle states as shown in the last equivalence in Eq.~(\ref{eq:4pointgreen}), provided the time arguments are exchanged at the same time. 

Depending on the ordering of the time arguments, the 4-point Green's function describes the propagation of either two-particle (pp), two-hole (hh) or particle-hole (ph) excitations. 

The Fourier transform of $K$ is defined as
\begin{multline}  
\bck{\alpha{\beta}}{K(\omega_{\alpha},\omega_{\beta},\omega_{\gamma},\omega_{\delta})}{\gamma{\delta}}=  \\
\int^{+\infty}_{-\infty}dt_{\alpha}dt_{\beta}dt_{\gamma}dt_{\delta}e^{i\omega_{\alpha}t_{\alpha}+i\omega_{2}t_{2}-i\omega_{{\gamma}}t_{{\gamma}}-i\omega_{{\delta}}t_{{\delta}}}\bck{\alpha{\beta}}{K(t_{\alpha},t_{\beta};t_{\gamma},t_{\delta})}{\gamma{\delta}},
\end{multline}
and the inverse relation as
\begin{multline}  
\bck{\alpha{\beta}}{K(t_{\alpha},t_{\beta};t_{\gamma},t_{\delta})}{\gamma{\delta}}= \\
\frac{1}{(2\pi)^{4}}\int^{+i\infty}_{-i\infty}d\omega_{\alpha}d\omega_{\beta}d\omega_{\gamma}d\omega_{\delta}e^{-i\omega_{\alpha}t_{\alpha}-i\omega_{\beta}t_{\beta}+i\omega_{\gamma}t_{\gamma}+i\omega_{\delta}t_{\delta}}\bck{\alpha{\beta}}{K(\omega_{\alpha},\omega_{\beta},\omega_{\gamma},\omega_{\delta})}{\gamma{\delta}}.
\end{multline}
When the Hamiltonian $\hat{H}$ is time-independent, $K$ does not depend on the sum of the time variables, meaning that it is independent of the variable $t=\frac{1}{4}(t_{\alpha}+t_{\beta}+t_{\gamma}+t_{\delta})$.

\begin{figure}[ht]
\centering
\includegraphics[width=5cm]{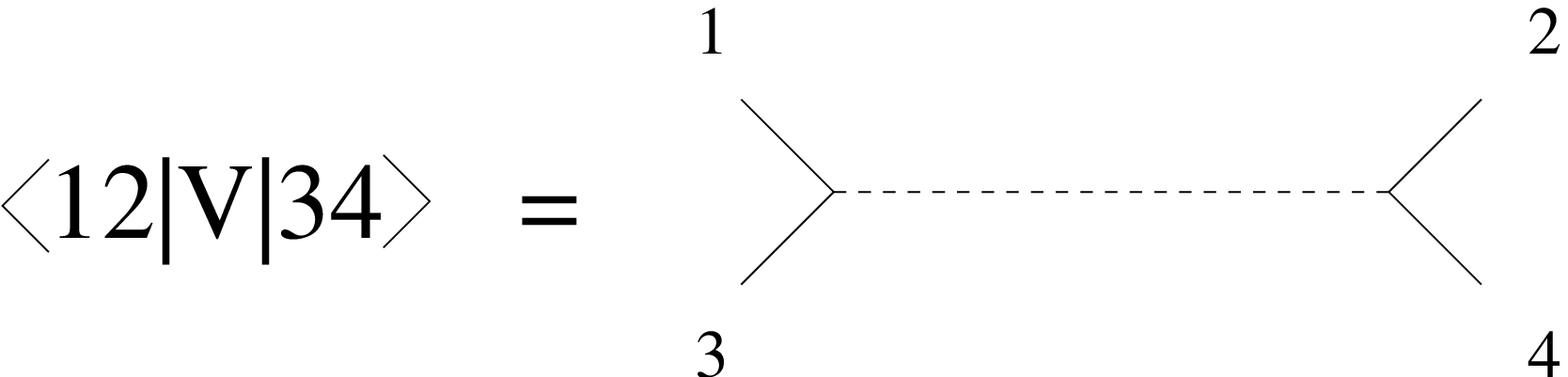}
\caption{The convention for numbering the legs of an interaction.} \label{fig:numbering}
\end{figure}

Before proceeding, we give a short summary of our Feynman diagram rules. The basic building blocks are interactions, represented by a horizontal line (dotted, wavy or thick-lined) connected by vertical particle/hole lines. The translation of the interaction lines are as matrix elements, the convention for numbering the incoming and outgoing states always being as shown in Fig.~\ref{fig:numbering}. 
Lines are associated with the full propagator, implying that particle/hole lines can be translated as either particle or hole propagators, each diagram giving rise to two Goldstone diagrams. A diagram with two parallel lines corresponds to two (skeleton) Goldstone diagrams, one with two particles and one with two holes propagating. A ring-like structure translates into a particle-hole pair. When combined, one of the two-particle/two-hole parallel lines might sometimes be part of a larger particle-hole bubble, and for aesthetic reasons no longer straight. No ambiguity should arise form this. Matrix elements of more complex operators than a simple interaction are represented as either rectangles (composite interactions) or circles (propagators), all conforming to the same numbering convention as the interaction matrix elements (Fig.~\ref{fig:numbering}). In some diagrams we use arrows on the internal lines. These are intended as a graphical means of showing which states are incoming and which are outgoing in the matrix elements of each operator, and thus have no separate physical translation.

\subsection{The interaction operator} \label{subsec:Gamma4pt}

\begin{figure}[ht]
\includegraphics[width=10cm]{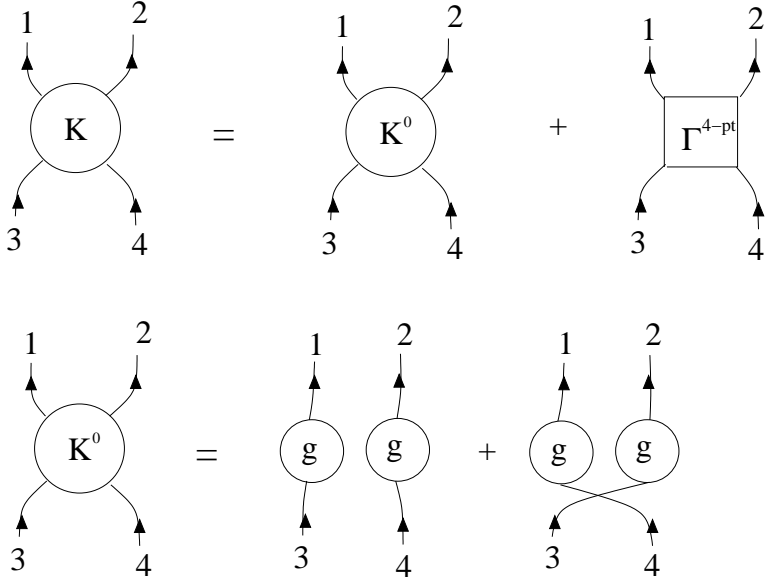}
\caption{The four-point Green's function $K$, separated into a set of unconnected diagrams $K^0$ and a set of connected diagrams. The unconnected diagrams can be summarized as consisting of two unconnected fully renormalized propagators and their exchange contributions, as shown in the lower equation. A line indicates the result of a Wick contraction, depending on the time ordering this could be either a particle or a hole. The arrows on the lines are meant to clarify the relationship between the matrix elements and the pictorial description, and do not distinguish particles from holes, any of the lines could be of either type. } \label{fig:fourptgreen}
\end{figure}

Using standard many-body perturbation techniques \cite{BlaizotRipka1986}, we can obtain a diagrammatic expansion for the four-point propagator in Eq.~(\ref{eq:4pointgreen}). We see that there are two classes of contributions, one unconnected class in which two different particle lines propagate without any interaction between each other, and a second group where the particle lines are connected by interaction lines. In Fig.~\ref{fig:fourptgreen} we sketch in a schematic way the two classes. 

The four-point interaction vertex $\Gamma^{\text{4-pt}}$ is called the interaction operator, defined as all two-line irreducible diagrams with fully renormalized propagators. To lowest order, $\Gamma^{\text{4-pt}}$ is identical to the two-body interaction $V$. To make the expressions a little more readable, we will henceforth use roman numerals on the incoming and outgoing states, and Greek letters for intermediate states. We can then express the diagram for $K$ given in Fig.~\ref{fig:fourptgreen} in terms of the four-point interaction vertex $\Gamma^{\text{4-pt}}$ as:
\begin{multline} \label{eq:4pointgreen+Gamma}
\bck{12}{K(t_1,t_2;t_3,t_4)}{34}=
i{}[g_{13}(t_1-t_3)g_{24}(t_2-t_4)-g_{14}(t_1-t_4)g_{23}(t_{2}-t_3)] \\
-\int{dt_{\alpha}}\int{dt_{\beta}}\int{dt_{\gamma}}\int{dt_{\gamma}}\sum_{\alpha{\beta}\gamma{\delta}}g_{1{\alpha}}(t_1-t_{\alpha})g_{2{\beta}}(t_{2}-t_{\beta}) \\
 \times \bck{\alpha{\beta}}{\Gamma^{\text{4-pt}}(t_{\alpha},t_{\beta};t_{\gamma},t_{\delta})}{\gamma{\delta}} g_{\gamma{3}}(t_{\gamma}-t_{3})g_{\delta{4}}(t_{\delta}-t_{4}).
\end{multline}
We define the Fourier transform of the interaction operator as 
\begin{multline}
\bck{\alpha{\beta}}{\Gamma^{\text{4-pt}}(\omega_{\alpha},\omega_{\beta},\omega_{\gamma},\omega_{\delta}}{\gamma{\delta}} \equiv \\
\int{dt_{\alpha}}\int{dt_{\beta}}\int{dt_{\gamma}}\int{dt_{\delta}} e^{i\omega_{\alpha}t_{\alpha}}e^{i\omega_{\beta}t_{\beta}}e^{-i\omega_{\gamma}t_{\gamma}}e^{-i\omega_{\delta}t_{\delta}}\bck{\alpha{\beta}}{\Gamma^{\text{4-pt}}(t_{\alpha},t_{\beta},t_{\gamma},t_{\delta})}{\gamma{\delta}}.
\end{multline}
If the bare interaction is time-independent, it does not depend on the energy, and consequently the interaction operator conserves energy and depends only on the incoming energies.

The Fourier transform of the four-point Green's function can be written as
\begin{multline} \label{eq:FTg4pt}
\bck{12}{K(\omega_1,\omega_{\beta},\omega_3,\omega_4)}{34}=2\pi{i}\delta(\omega_1+\omega_2-\omega_3-\omega_4)\\
\times \bigl{[}2\pi{}\delta(\omega_1-\omega_3)g_{13}(\omega_1)g_{24}(\omega_2)-2\pi{}\delta(\omega_1-\omega_4)g_{14}(\omega_1)g_{23}(\omega_2)\bigr{]}\\
-\sum_{\alpha{\beta}\gamma{\delta}}g_{1{\alpha}}(\omega_1)g_{2{\beta}}(\omega_2)\bck{\alpha{\beta}}{\Gamma^{\text{4-pt}}(\omega_{\alpha},\omega_{\beta},\omega_{\gamma},\omega_{\delta})}{\gamma{\delta}}g_{\gamma{3}}(\omega_3)g_{\delta{4}}(\omega_4).
\end{multline}

We call the first part of the four-point propagator (the first term in Eq.~(\ref{eq:4pointgreen+Gamma})) for the non-interacting or free four-point propagator. It consists of a product of two one-particle propagators. From the above expression we see that the Fourier transform of the non-interacting four-point propagator $K^{0}$ is given by:
\begin{multline}
\bck{12}{K^0(\omega_1,\omega_2,\omega_3,\omega_4)}{34} = 2\pi{i}\delta(\omega_1+\omega_2-\omega_3-\omega_4)\\
\times \bigl{[}2\pi{}\delta(\omega_1-\omega_3)g_{13}(\omega_1)g_{24}(\omega_2)-2\pi{}\delta(\omega_1-\omega_4)g_{14}(\omega_1)g_{23}(\omega_2)\bigr{]}.
\end{multline}

\subsection{Self energy}\label{subsec:selfenergy}

\begin{figure}[ht]
\includegraphics[width=10cm]{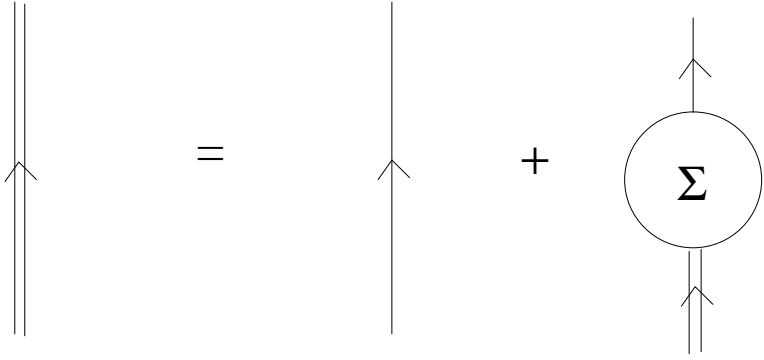}
\caption{The diagrammatic representation of the Dyson equation. The single line represent the unperturbed propagator, the double line represents the full (dressed) propagator.} \label{fig:dyson}
\end{figure}

To find an expression for the one-particle propagator, we once again use standard many-body perturbation techniques \cite{Dickhoff2005,FetterWalecka1971,BlaizotRipka1986}. This gives the Dyson equation, giving a decomposition of the propagator in terms of the irreducible self energy $\Sigma$ (also called the proper self energy or the mass operator): 
\begin{equation} \label{eq:dyson}
g_{\alpha{\beta}}(\omega) = g^{0}_{\alpha{\beta}}(\omega)+\sum_{\gamma{\delta}}g^{0}_{\alpha{\gamma}}(\omega)\Sigma(\gamma,{\delta};\omega)g_{\delta{\beta}}(\omega).
\end{equation}
The self energy is the one-line irreducible diagrammatic insertions to the one-particle propagator, as shown in the diagrammatic representation of the Dyson equation in Fig.~\ref{fig:dyson}. By iterating on this we generate the exact one-particle propagator, provided the exact irreducible self-energy can be found. This is unfortunately in general not possible.

In terms of diagrams, a one-particle propagator including self-energy insertions is called a dressed propagator, often drawn as a double line. In the case of our Parquet diagrams, however, all propagators are dressed, and for the sake of simplicity, we have chosen to draw them as single lines nonetheless. The only exception being in Fig.~\ref{fig:dyson}, where the single line represents the unperturbed propagator.

\begin{figure}[ht]
\includegraphics[width=13cm]{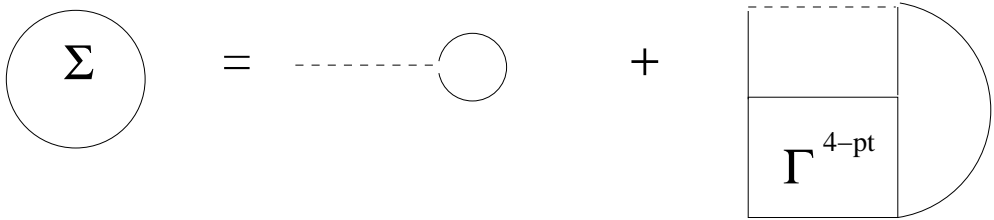}
\caption{The self energy $\Sigma$ expressed by the interaction operator $\Gamma^{\text{4-pt}}$. The propagators are dressed propagators.} \label{fig:sigma}
\end{figure}

We can find a useful relation between the self energy and the interaction operator $\Gamma^{\text{4-pt}}$ from the equation of motion of the one-particle propagator given in Eq.~(\ref{eq:eq_mot_1prop}). Inserting the expression in Eq.~(\ref{eq:4pointgreen+Gamma}) for the 4-point propagator, we obtain:
\begin{multline}
i\frac{\partial}{\partial{t}}g_{12}(t-{t}^\prime)=\delta(t-t^{\prime})\delta_{12} +\epsilon_{1}g_{12}(t-{t}^\prime)
-i{}\sum_{\alpha{\beta}\gamma}\bck{1{\alpha}}{V}{\beta{\gamma}}g_{\beta{\alpha}}(t-t^{+})g_{\gamma{2}}(t-t^{\prime})\\
+ \frac{1}{2}\sum_{\alpha{\beta}\gamma}\sum_{\delta{\xi}\mu{\nu}}\int{d}t_{\delta}\int{d}t_{\xi}\int{d}t_{\mu}\int{d}t_{\nu} g_{\gamma{\delta}}(t-t_{\delta})g_{\beta{\xi}}(t-t_{\xi})g_{\nu{\alpha}}(t_{\nu}-t) \\
\times \bck{\delta{\xi}}{\Gamma^{\text{4-pt}}(t_{\delta},t_{\xi},t_{\mu},t_{\nu})}{\mu{\gamma}}g_{\mu{2}}(t_{\mu}-t^{\prime}).\\
\end{multline}
Taking the Fourier transform of the above expression and performing some algebra, see Ref.~\cite{Dickhoff2005}, we arrive at an expression for the one-particle propagator which is identical to the Dyson equation, provided we make the identification
\begin{multline} \label{eq:Sigma(E)}
\Sigma(1,2;\omega)= -i \int_{C\uparrow}\frac{d\omega_1}{2\pi}\sum_{\alpha{\beta}}\bck{1{\alpha}}{V}{2{\beta}}g_{\alpha{\beta}}(\omega_1)\\
+ \frac{1}{2}\int{\frac{d\omega_1}{2\pi}}\int{\frac{d\omega_2}{2\pi}}\sum_{\alpha{\beta}\gamma{\delta}\mu{\nu}}\bck{1{\alpha}}{V}{\beta{\gamma}}g_{\beta{\delta}}(\omega_1)g_{\gamma{\mu}}(\omega_2) \\
\times \bck{\delta{\mu}}{\Gamma^{\text{4-pt}}(\omega_1,\omega_2,\omega,\omega_1+\omega_2-\omega)}{2{\nu}}g_{\nu{\alpha}}(\omega_1+\omega_2-\omega).
\end{multline}
Here the integral in the first expression is a contour integral along the real axis to be closed in the upper half plane, as indicated by the $C\uparrow$ subscript. The expression for $\Sigma$ is shown diagrammatically in Fig.~\ref{fig:sigma}.
Equations (\ref{eq:FTg4pt}) and (\ref{eq:Sigma(E)}) together with the Dyson Eq.~(\ref{eq:dyson}) give the exact description of the one-particle propagator if the single particle propagators in the interaction operator are dressed, that is, the self energy insertions are included. This gives a set of non-linear equations, and any solution procedure needs to include some sort of self-consistency scheme, as will be further discussed in Sections \ref{subsec:selfparquet} and \ref{sec:itersolution}.

\subsection{The eigenvalue equation method} \label{subsec:sp_matinv}

We can write the Dyson equation, see again Eq.~(\ref{eq:dyson}), as a matrix equation using the notation $[g]$ as shorthand for the matrix with $g_{\alpha{\beta}}$ with indices $\alpha, \beta$. Assuming the unperturbed propagator to be diagonal, with the inverse given as $[g^0]^{-1}=\omega-[e]$, the Dyson equation can be generically written as
\begin{equation} \label{eq:g^-1}
[g(\omega)] = [\omega \cdot{\bf 1} -([e]+[\Sigma(\omega)])]^{-1},
\end{equation}
where $e_{\alpha}$ represents the energies of the unperturbed Hamiltonian, $[e]$ being a diagonal matrix with $e_{\alpha}$ at the diagonal entries. From this we see that the poles of the propagator are the roots $\omega_{\lambda}$ of the equation
\be\label{eq:epsilon(omega)}
([e]+[\Sigma(\omega_{\lambda})])\ket{\lambda}=\omega_{\lambda}\ket{\lambda}.
\ee
Recalling the Lehmann representation of the propagator, see Eq.~(\ref{eq:1p_prop_omega}), we can identify these roots as the energies of the $N\pm{1}$ systems. The residue matrix $[S_{\lambda}]$ of the propagator at the pole $\omega_{\lambda}$ is given by \cite{Dickhoff2005}:
\begin{equation}
[S_{\lambda}]=\lim_{\omega{\to}\omega_{\lambda}}(\omega-\omega_{\lambda})[g(\omega)]= 
\frac{1}{1-\bra{\tilde{\lambda}}[\Sigma^{\prime}(\omega_{\lambda})]\ket{\lambda}}\ket{\lambda}\bra{\tilde{\lambda}}\\
= s_{\lambda}\ket{\lambda}\bra{\tilde{\lambda}}.
\end{equation}
The eigenstate $\bra{\tilde{\lambda}}$ is the corresponding left eigenstate of the operator in Eq.~(\ref{eq:epsilon(omega)}). The left and right eigenstates are assumed to be normalized according to
\begin{equation} \label{eq:lambdanorm}
\bk{\tilde{\lambda}}{\lambda} = 1.
\end{equation}
We assume that the propagator has only simple poles, the expression for the degenerate case is somewhat more involved. Now we can write the propagator as
\be
[g(\omega)]=\sum_{\lambda}\frac{[S_{\lambda}]}{\omega-\omega_{\lambda}}.
\ee
The eigenvalue Eq.~(\ref{eq:epsilon(omega)}) is more complicated than an ordinary eigenvalue equation, as $\Sigma$ depends on the energy and has to be calculated at the unknown eigenvalue. There is no longer only one solution, but a set of different $\omega_{\lambda}$ which can be quite large (depending on the number of poles in $\Sigma$). Physically, this means that adding or removing one particle from the ground state no longer leaves the system in one definite state, rather there are several possible states, each with its own amplitude. The sum over these are still unity. The independent-particle model is no longer appropriate, however, calculations show that at least for nuclear system, much of the single-particle strength is still concentrated in a single state for states close to the Fermi energy. States further away, either deep down in the nuclear well or high up, closer to the continuum, get smeared out and cannot properly be called single-particle states any more. 

From the one-body propagator it is possible to find the energy of the ground state by using the so-called  Migdal-Galitski-Koltun sum rule \cite{Dickhoff2005}(see also Boffi in Ref.~\cite{boffi1971}):
\begin{equation} \label{eq:MGKfin}
\begin{split}
 E_{0}^{A} &  = \langle{\Psi_{0}^{A}|\hat{H}|\Psi_{0}^{A}\rangle} \\
& = \frac{1}{2} \sum_{\alpha{\beta}}\bck{\alpha}{T}{\beta}\sum_{\lambda<\lambda_{F}}S_{\lambda}\bk{\alpha}{\lambda}\bk{\tilde{\lambda}}{\beta} +  \frac{1}{2} \sum_{\alpha{\beta}}\sum_{\lambda<\lambda_{F}}S_{\lambda}\bk{\alpha}{\lambda}\bk{\tilde{\lambda}}{\beta}\omega_{\lambda}.
\end{split}
\end{equation}

\section{Parquet theory} \label{sec:parquet}

The formalism presented in the previous section requires calculations of several infinite sums, and thus we need some procedure to handle these. The Parquet method offers an approximation to the interaction operator $\Gamma^{\text{4-pt}}$ which includes a large, infinite subset of the full set of diagrams. 

 We observe that there are several different possibilities for reducing this four-time operator down to a two-time operator. Depending on the physical system in question, reductions to a ladder or a ring operator has been used to include either pphh (which include for example 2p2h excitations, see discussion below) or ph correlations respectively. However, as argued by Jackson and Wettig \cite{Jackson1994}, neither of these approaches meet some basic requirements of a many-body theory to be certain of convergence. These authors have further argued that a necessary (but perhaps not sufficient) condition for any many-body summation of diagram to converge, is that both pp and hh ladders and ph chains be summed to all orders. 

A Green's function based like the Faddeev random-phase approximation of Barbieri and co-workers, see for example \cite{Barbieri2006,Dickhoff2004}, couples ladder diagrams and ring diagrams to all orders for the self-energy. The parquet method offers a method of doing this in a fairly straightforward manner and includes more complicated 2p2h correlations.  However, in order to perform such calculations, approximations are necessary, in particular with regard to the treatment of the energy dependence of the propagators. These approximations are discussed below.

Thus we first discuss the principle behind the Parquet theory, namely the different channels in which iterative expressions for the interaction operator can be found (Section \ref{subsec:channels}). Then we need some more notation, given in Section \ref{subsec:angmom}, before we are ready to discuss the possible two-time reductions of the four-point propagators in Section \ref{subsec:twotimeprop}. In Section \ref{subsec:selfparquet} the Parquet equations are given in a form suitable as starting point for numerical implementation.  

\subsection{Channels: Equivalent ways of building the Interaction Operator} \label{subsec:channels}

\begin{figure}[ht]
\center
\includegraphics[width=13cm]{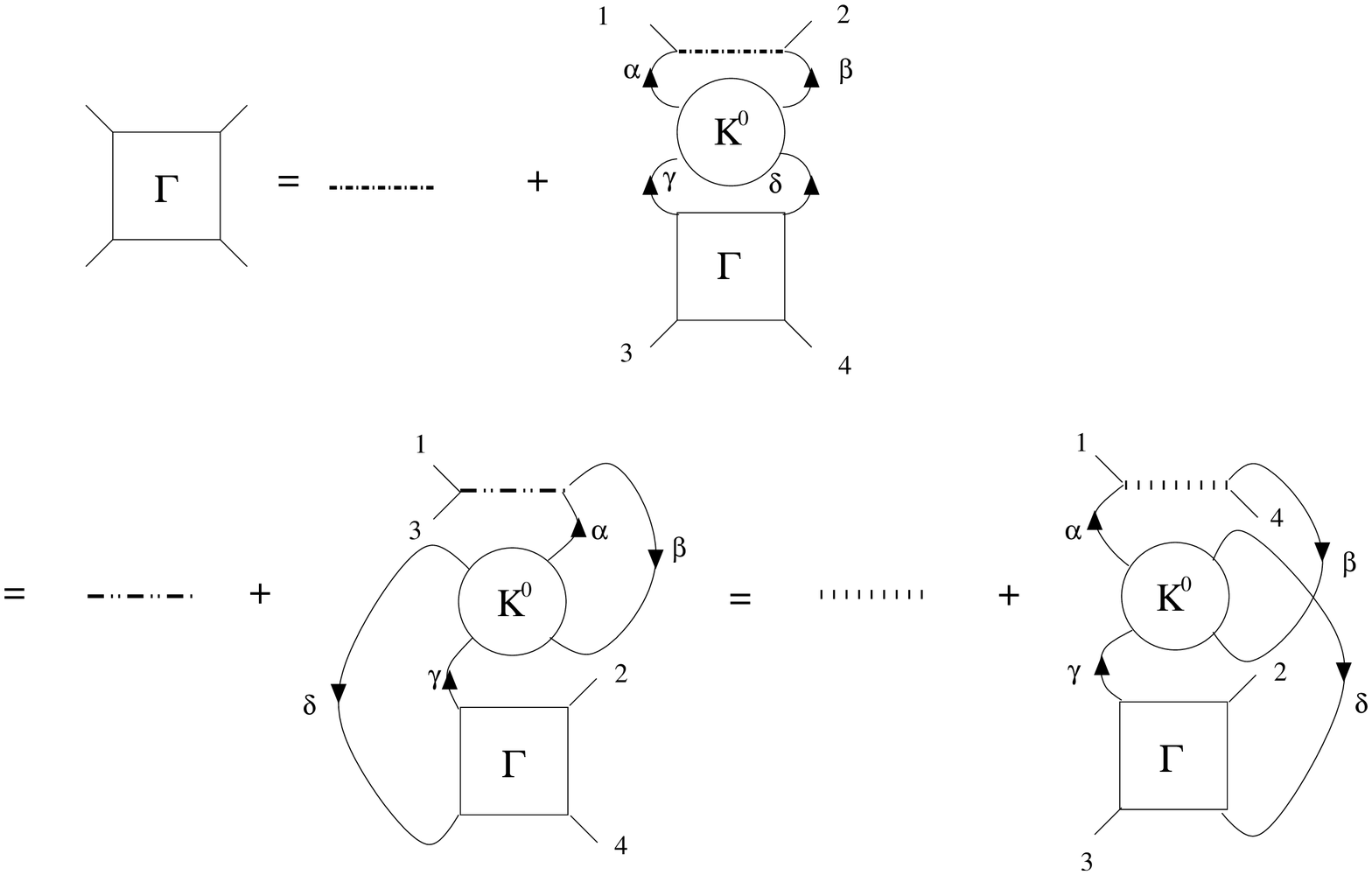}
\caption{Iterative expressions for $\Gamma^{\text{4-pt}}$ in the [12]-channel, the [13]-channel and the [14]-channel. The internal arrows determine which states are incoming and which are outgoing in the matrix elements of each operator.} \label{fig:Gamma1234}
\end{figure}

We can obtain iterative expressions for the interaction operator defined in Section \ref{subsec:Gamma4pt} by examining it order by order. To first order, it is just the bare interaction. The next order consists of two bare interactions connected by the non-interacting propagator $K^{0}$, third order is found by connecting a third interaction by another $K^{0}$, and so on. There are three equivalent ways of connecting the legs of the interactions, as shown in Fig.~\ref{fig:Gamma1234}. We name the different possibilities according to the numbering shown in Fig.~\ref{fig:numbering}. Thus, if we connect the lines 1 and 2 we are in the $[12]$ channel or particle-particle channel, while connecting the lines 1 and 3 or the lines 1 and 4 give the $[13]$ channel or the $[14]$ channel, respectively. These two latter channels are called the particle-hole channels. The $[12]$, $[13]$ and $[14]$ channels are the equivalents of the Mandelstam variables $s,t$ and $u$ from relativistic quantum mechanics~\cite{BlaizotRipka1986}.

A diagram contributing to the interaction operator $\Gamma^{\text{4-pt}}$ either can or cannot be split into two disconnected parts, one containing the legs 1 and 2 and the other the legs 3 and 4 by cutting two internal lines. If this splitting is impossible, the diagram is said to be simple in the [12] channel, if it is possible, the diagram is called non-simple. The particle-particle interaction $\mathcal{V}^{12}$ is defined as the sum over all the [12]-simple diagrams.
\begin{figure}[ht]
\center
\includegraphics[width=13cm]{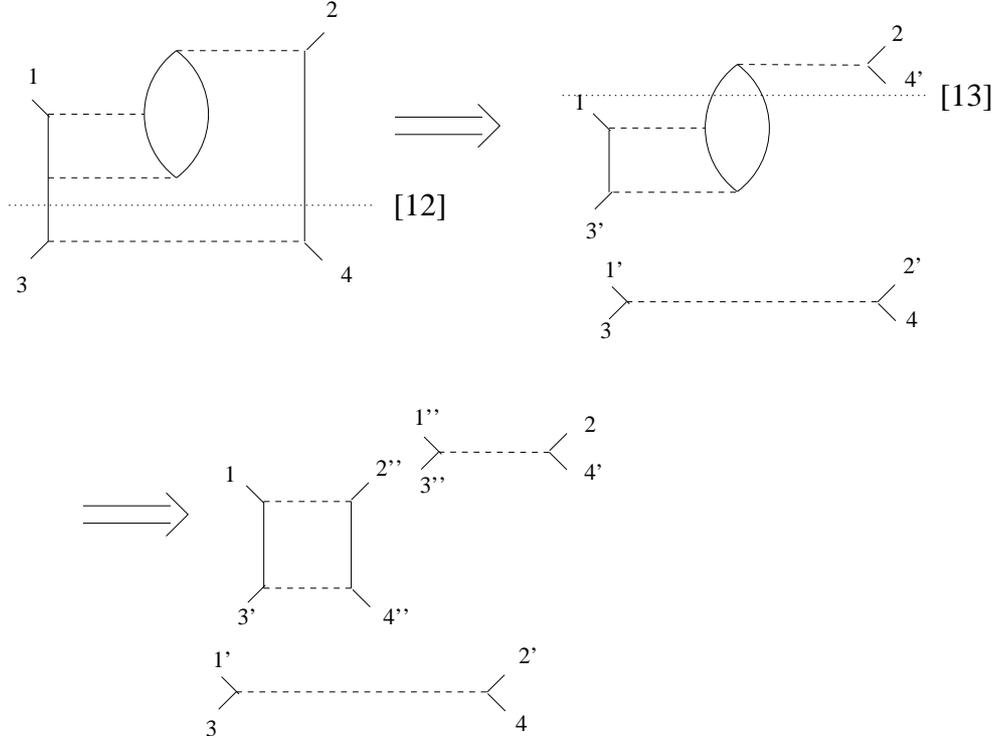}
\caption{Example diagram broken down into components. The original diagram is non-simple in the [12] channel and can be split into two parts by cutting two internal lines in such a manner that one part contains legs 1 and 2 and the other legs 3 and 4. The results are both simple in the [12] channel. The upper part is non-simple in the [13] channel, and can be split into two parts, one containing legs 1 and $3^{\prime}$, the other 2 and $4^{\prime}$. Both these are simple in the [12] channel. The final composite diagram is non-simple in the [12] channel.} \label{fig:channelex}
\end{figure}
It is easily seen that the full interaction operator $\Gamma^{\text{4-pt}}$ is obtained by iterating over $\mathcal{V}^{12}$, as shown in the first line of Fig.~\ref{fig:Gamma1234}, where the dash-dot line represents an [12]-simple interaction.  The equation for the vertex translates into the well-known Bethe-Salpeter equation:
 \begin{multline}\label{eq:Gamma12}
\bck{12}{\Gamma^{\text{4-pt}}(\omega_1,\omega_{\beta},\omega_3,\omega_4)}{34}= \bck{12}{\mathcal{V}^{12}(\omega_1,\omega_2,\omega_3,\omega_4)}{34} \\
+ \frac{1}{2}\int{\frac{d\omega_{\alpha}}{2\pi{}}}\int{\frac{d\omega_{\beta}}{2\pi{}}}\int{\frac{d\omega_{\gamma}}{2\pi{}}}\int{\frac{d\omega_{\delta}}{2\pi{}}}\sum_{\alpha{\beta}\gamma{\delta}}\bck{12}{\mathcal{V}^{12}(\omega_1,\omega_2,\omega_{\alpha},\omega_{\beta})}{\alpha{\beta}} \\
\times \bck{\alpha{\beta}}{K^{0}(\omega_{\alpha},\omega_{\beta},\omega_{\gamma},\omega_{\delta})}{\gamma{\delta}}\bck{\gamma{\delta}}{\Gamma^{\text{4-pt}}(\omega_{\gamma},\omega_{\delta},\omega_3,\omega_4)}{34}.
\end{multline}
The factor $\frac{1}{2}$ stems from the symmetry of the interaction with respect to the exchange of indices.

Similarly, we define the particle-hole interaction $\mathcal{V}^{13}$ as the sum over all diagrams which are simple in the [13] channel, that is, all diagrams that cannot be split into one part containing the external lines 1 and 3, and another containing the lines 2 and 4. The particle-hole interaction $\mathcal{V}^{14}$ is defined as the sum over all [14]-simple diagrams (diagrams which cannot be split into one part containing the external lines 1 and 4, and another containing the lines 2 and 3). An example diagram and the splitting of different components is shown in Fig.~\ref{fig:channelex}.
 
\begin{figure}[ht]
\center
\includegraphics[width=10cm]{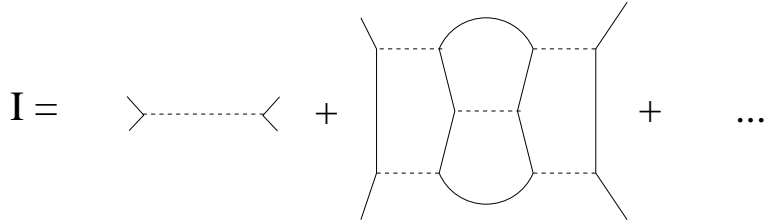}
\caption{The diagram class I, class of diagrams simple in all three channels. The first diagram other than the bare interaction is of fifth order in the interaction. We have included only the first contribution (the bare interaction) in our calculations.} \label{fig:butter}
\end{figure}

Each of these will give the full interaction operator if we iterate as shown in Fig.~\ref{fig:Gamma1234}, where the dash-dot-dot (large dot) line represents an [13]-simple ([14]-simple) interaction. The Bethe-Salpeter equations corresponding to these diagrams are 
\begin{multline}\label{eq:Gamma13}
\bck{12}{\Gamma^{\text{4-pt}}(\omega_1,\omega_{2},\omega_3,\omega_4)}{34}= \bck{12}{\mathcal{V}^{13}(\omega_1,\omega_2,\omega_3,\omega_4)}{34} \\
+ \int{\frac{d\omega_{\alpha}}{2\pi{}}}\int{\frac{d\omega_{\beta}}{2\pi{}}}\int{\frac{d\omega_{\gamma}}{2\pi{}}}\int{\frac{d\omega_{\delta}}{2\pi{}}}\sum_{\alpha{\beta}\gamma{\delta}}\bck{1\beta}{\mathcal{V}^{13}(\omega_1,\omega_{\beta},\omega_{3},\omega_{\alpha})}{3{\alpha}} \\
\times \bck{\delta{\alpha}}{K^{0}(\omega_{\delta},\omega_{\alpha},\omega_{\gamma},\omega_{\beta})}{\gamma{\beta}}\bck{{\gamma}2}{\Gamma^{\text{4-pt}}(\omega_{\gamma},\omega_{2},\omega_{\delta},\omega_4)}{\delta{4}},
\end{multline}
and
\begin{multline}\label{eq:Gamma14}
\bck{12}{\Gamma^{\text{4-pt}}(\omega_1,\omega_{\beta},\omega_3,\omega_4)}{34}= \bck{12}{\mathcal{V}^{14}(\omega_1,\omega_2,\omega_3,\omega_4)}{34} \\
+ \int{\frac{d\omega_{\alpha}}{2\pi{}}}\int{\frac{d\omega_{\beta}}{2\pi{}}}\int{\frac{d\omega_{\gamma}}{2\pi{}}}\int{\frac{d\omega_{\delta}}{2\pi{}}}\sum_{\alpha{\beta}\gamma{\delta}}\bck{1\beta}{\mathcal{V}^{14}(\omega_1,\omega_{\beta},\omega_{\alpha},\omega_{4})}{\alpha{4}} \\
\times \bck{{\alpha}\delta}{K^{0}(\omega_{\alpha},\omega_{\delta},\omega_{\gamma},\omega_{\beta})}{\gamma{\beta}}\bck{{\gamma}2}{\Gamma^{\text{4-pt}}(\omega_{\gamma},\omega_{2},\omega_{3},\omega_{\delta})}{3{\delta}}.
\end{multline}
Combining the information in these three equations, we see that the diagrams of $\Gamma^{\text{4-pt}}$ fall into four classes. One class of diagrams consists of diagrams that are simple in any of the three channels, that is, these diagrams cannot be cut into two separate pieces by cutting any two lines. The lowest-order member of this class is the bare interaction, and the next is of fifth order in the bare interaction, shown in Fig.~\ref{fig:butter}. We call this class $I$. In our calculations we will only include the first term in this series, that is, $I=V$.

Then there is the class of diagrams which are non-simple in the [12] channel, generated by repeated iterations of the type shown in Eq.~(\ref{eq:Gamma12}). We call this class the $L$ diagrams. The conventional ladder diagrams is a subset of this class. In terms of the $\mathcal{V}^{12}$ interaction, we have that
\begin{multline} \label{eq:pqL4ptV} 
\bck{12}{L(\omega_1,\omega_{\beta},\omega_3,\omega_4)}{34} = \\
\int{\frac{d\omega_{\alpha}}{2\pi{}}}\int{\frac{d\omega_{\beta}}{2\pi{}}}\int{\frac{d\omega_{\gamma}}{2\pi{}}}\int{\frac{d\omega_{\delta}}{2\pi{}}} \sum_{\alpha{\beta}\gamma{\delta}}\bck{12}{\mathcal{V}^{12}(\omega_1,\omega_2,\omega_{\alpha},\omega_{\beta})}{\alpha{\beta}} \\
\times \bck{\alpha{\beta}}{K^{0}(\omega_{\alpha},\omega_{\beta},\omega_{\gamma},\omega_{\delta})}{\gamma{\delta}}\bck{\gamma{\delta}}{\Gamma^{\text{4-pt}}(\omega_{\gamma},\omega_{\delta},\omega_3,\omega_4)}{34}.
\end{multline}
Then there are the classes which are made from iterations of the types in equations (\ref{eq:Gamma13}) and (\ref{eq:Gamma14}). It is fairly easy to show that each of the diagrams in the [14] channel class have an exchange counterpart in the [13] channel class. Working with antisymmetrized matrix elements in either channel will thus include all diagrams, and only one of the channels need be included in the calculation. We have chosen to work with the [13] channel class, and call this the $R$ class. The diagrams summed in a standard RPA calculation result in a subset of this class. Expressing $R$ in terms of $\mathcal{V}^{13}$ gives
\begin{multline} \label{eq:pqR4ptV}
\bck{12}{R(\omega_1,\omega_{\beta},\omega_3,\omega_4)}{34} = \\
\int{\frac{d\omega_{\alpha}}{2\pi{}}}\int{\frac{d\omega_{\beta}}{2\pi{}}}\int{\frac{d\omega_{\gamma}}{2\pi{}}}\int{\frac{d\omega_{\delta}}{2\pi{}}} \sum_{\alpha{\beta}\gamma{\delta}}\bck{1\alpha}{\mathcal{V}^{13}(\omega_1,\omega_2,\omega_{\alpha},\omega_{\beta})}{3{\beta}} \\
\times \bck{\delta{\alpha}}{K^{0}(\omega_{\delta},\omega_{\alpha},\omega_{\gamma},\omega_{\beta})}{\gamma{\beta}}\bck{{\gamma}2}{\Gamma^{\text{4-pt}}(\omega_{\gamma},\omega_2,\omega_{\delta},\omega_4)}{\delta{4}}.
\end{multline}
With these definitions we see that the [12]-simple interaction $\mathcal{V}^{12}$ can be written as the sum $I+R$ and that the [13]-simple interaction $\mathcal{V}^{13}$ is the sum $I+L$. The total interaction operator can be written as the sum of all three diagram classes: 
\begin{multline} \label{eq:pqGamma4pt}
\bck{12}{\Gamma^{\text{4-pt}}(\omega_1,\omega_2,\omega_3,\omega_4)}{34} = \bck{12}{I(\omega_1,\omega_2,\omega_3,\omega_4)}{34}\\
 + \bck{12}{L(\omega_1,\omega_2,\omega_3,\omega_4)}{34} + \bck{12}{R(\omega_1,\omega_2,\omega_3,\omega_4)}{34} \\ \equiv \bck{12}{(I + L + R)(\omega_1,\omega_2,\omega_3,\omega_4)}{34}.
\end{multline}
Rewriting Eqs.~(\ref{eq:pqR4ptV}) and (\ref{eq:pqGamma4pt}) in terms of $I$, $L$ and $R$ we obtain
\begin{multline} \label{eq:pqL4pt} 
\bck{12}{L(\omega_1,\omega_2,\omega_3,\omega_4)}{34} = \\
\int{\frac{d\omega_{\alpha}}{2\pi{}}}\int{\frac{d\omega_{\beta}}{2\pi{}}}\int{\frac{d\omega_{\gamma}}{2\pi{}}}\int{\frac{d\omega_{\delta}}{2\pi{}}} \sum_{\alpha{\beta}\gamma{\delta}}\bck{12}{(I+R)(\omega_1,\omega_2,\omega_{\alpha},\omega_{\beta})}{\alpha{\beta}} \\
\times \bck{\alpha{\beta}}{K^{0}(\omega_{\alpha},\omega_{\beta},\omega_{\gamma},\omega_{\delta})}{\gamma{\delta}}(\bck{\gamma{\delta}}{(I +L+R )(\omega_{\gamma},\omega_{\delta},\omega_3,\omega_4)}{34}.
\end{multline}
and
\begin{multline} \label{eq:pqR4pt}
\bck{12}{R(\omega_1,\omega_2,\omega_3,\omega_4)}{34} = \\
\int{\frac{d\omega_{\alpha}}{2\pi{}}}\int{\frac{d\omega_{\beta}}{2\pi{}}}\int{\frac{d\omega_{\gamma}}{2\pi{}}}\int{\frac{d\omega_{\delta}}{2\pi{}}} \sum_{\alpha{\beta}\gamma{\delta}}\bck{1\alpha}{(I+L)(\omega_1,\omega_2,\omega_{\alpha},\omega_{\beta})}{3{\beta}} \\
\times \bck{{\beta}\gamma}{K^{0}(\omega_{\alpha},\omega_{\beta},\omega_{\gamma},\omega_{\delta})}{\alpha{\delta}}\bck{\delta{2}}{(I+L+R)(\omega_{\gamma},\omega_{\delta},\omega_3,\omega_4)}{{\gamma}4}.
\end{multline}
The three equations (\ref{eq:pqGamma4pt}), (\ref{eq:pqL4pt}) and (\ref{eq:pqR4pt}) together constitute the Parquet equations. In addition to these, some scheme for treating the self energy consistently has to be made. This is discussed in Section \ref{subsec:selfparquet}. Furthermore, some simplifications with respect to the energy dependence is needed, as will be discussed below. However, before doing that, we need to introduce some additional notation.

\subsection{Angular momentum recoupling}\label{subsec:angmom}

To reduce the basis space, it is convenient to introduce an angular-momentum coupled basis, that is, a basis where two single particle states are coupled to total angular momentum $J$. Each channel naturally give rise to its own coupling scheme. When necessary we indicate the coupling in the matrix elements as $\clbck{1}{2}{V}{3}{4}{}_J$ to signify coupling between states 1 and 2 to total angular momentum $J$ and $M$. Other quantum numbers such as for example the total isospin projection
can easily be added.
In our harmonic oscillator single-particle basis, the two-particle basis states are independent of $M$, giving a huge reduction in computational complexity. The conventional coupling order in the particle-hole channels is to couple incoming state to outgoing state. This is the same notation as in Ref.~\cite{Kuo1981}, see this work for an extensive discussion and examples. 

The $[12]$ coupling scheme is the standard coupling scheme. The antisymmetrized two-particle $J_{[12]}$-coupled state is given by
\begin{multline}
 \cket{1}{2} = \ket{(n_1j_1l_1s_1t_{z1})(n_2j_2l_2s_2t_{z2})JM} = \\
\frac{1}{\sqrt{2}}(1-(-1))^{j_1+j_2-J}\sum_{m_1m_2}\sum_{{m_l}_1{m_l}_2}\sum_{{m_s}_1{m_s}_2}
\bk{j_1m_1j_2m_2}{JM}\bk{l_1{m_l}_1s_1{m_s}_1}{j_1m_1} \\
\bk{l_2{m_l}_2s_2{m_s}_2}{j_2m_2}\quad \ket{n_1j_1m_1l_1s_1t_{z1}} \otimes \ket{n_2j_2m_2l_2s_2t_{z2}},
\end{multline}
where the $\bk{l_1{m_l}_1s_1{m_s}_1}{j_1m_1}$ are Clebsch-Gordan coefficients.

Angular momentum algebra gives the following relations between matrix elements with different coupling schemes (see for example Ref.~\cite{Kuo1981})
\begin{equation}
\crbck{1}{2}{V}{3}{4}{}_{J} = \sum_{J^{\prime}} (-)^{j_1+j_4+J+J^{\prime}}\hat{J^{\prime}}^2 \left\{
      \begin{array}{ccc}
       j_3&j_1&J\\j_2&j_4&J'
      \end{array}
       \right\} \clbck{1}{2}{V}{3}{4}{}_{J^{\prime}}
\end{equation}
and 
\begin{equation}
\clbck{1}{2}{V}{3}{4}{}_{J} = \sum_{J^{\prime}} (-)^{j_1+j_4+J+J^{\prime}}\hat{J^{\prime}}^2 \left\{
      \begin{array}{ccc}
       j_3&j_1&J\\j_2&j_4&J'
      \end{array}
       \right\} \crbck{1}{2}{V}{3}{4}{}_{J^{\prime}},
\end{equation}
where $\hat{J}=\sqrt{2J+1}$. Similar relations hold between the [14] channel and the other two channels.

We have also found it useful to employ a matrix notation in the [13] channel that allows us to formulate the Parquet equations as matrix equations. We define the $J_{[13]}$-coupled matrix element as 
\begin{equation} \label{eq:<13>}
\bck{\widehat{13}}{V}{\widehat{24}}{}_{J} \equiv  \crbck{1}{2}{V}{3}{4}{}_{J}
\end{equation}
Employing the $J_{[13]}$-coupled matrix elements, the equations in the [13]-channel can be rewritten to a form close to matrix equations. The equation for $R$, see Eq.~(\ref{eq:pqR4pt}) becomes
\begin{multline} \label{eq:pqR4pt_13}
\bck{12}{R(\omega_1,\omega_2,\omega_3,\omega_4)}{34} = \bck{\widehat{13}}{R(\omega_1,\omega_2,\omega_3,\omega_4)}{\widehat{24}} = \\
\int{\frac{d\omega_{\alpha}}{2\pi{}}}\int{\frac{d\omega_{\beta}}{2\pi{}}}\int{\frac{d\omega_{\gamma}}{2\pi{}}}\int{\frac{d\omega_{\delta}}{2\pi{}}} \sum_{\alpha{\beta}\gamma{\delta}}(\bck{\widehat{13}}{(I+L)(\omega_1,\omega_2,\omega_{\alpha},\omega_{\beta})}{\widehat{{\beta}\alpha}} \\
\times \bck{\widehat{{\delta}\gamma}}{K^{0}(\omega_{\delta},\omega_{\alpha},\omega_{\gamma},\omega_{\beta})}{\widehat{\alpha{\beta}}}\bck{\widehat{{\gamma}\delta}}{(I+L+R)(\omega_{\gamma},\omega_{2},\omega_{\delta},\omega_4)}{\widehat{{2}4}}.
\end{multline}
This expression is easily transformed into a matrix equation by a suitable transformation of the $K^0$ matrix. The expressions in the [12]-channel lend themselves to such a formulation immediately with $J_{[12]}$-coupled matrix elements.

\subsection{Two-time propagators}\label{subsec:twotimeprop}

The four-time four-point Green's function can be reduced to a two-time operator by either requiring the ``upper'' and ``lower'' times of the diagram be pairwise equal ($t_3=t_4$ and $t_1=t_2$), or by setting the ``left-hand'' times and the ``right-hand'' times of the diagram equal, that is, $t_3=t_1$ and  $t_4=t_2$. The first choice gives rise to the ladder reduction, while the second is the basis for the RPA approach, and we call it the ring (or chain) reduction. 

The reduction from the four-point Green's function to different two-time operators corresponds to the reductions of the Bethe-Salpeter equation for relativistic spinors to different non-relativistic equations as the Lippmann-Schwinger-equation~\cite{Sakurai1994}.

\subsubsection{The ladder propagator} 
The ladder two-time reduction of the four-point Green's function is defined by
\begin{equation}\label{eq:Gpphh_4pt}
\begin{split}
\bck{12}{\mathcal{G}^{pphh}(t-t^{\prime})}{34}{\equiv}&\lim_{t_2\rightarrow{t}}\lim_{t_4\rightarrow{t^{\prime}}}\bck{12}{K(t,t_2,t^{\prime},t_4)}{34}\\
=&\lim_{t_2\rightarrow{t}}\lim_{t_4\rightarrow{t^{\prime}}}-i\bck{\Psi^{N}_0}{\mathcal{T}[c_{2_H}(t_2)c_{1_H}(t)c^{\dagger}_{3_H}(t^{\prime})c^{\dagger}_{4_H}(t_4)]}{\Psi^{N}_0} \\
=&i[g_{13}(t-t^{\prime})g_{24}(t-t^{\prime}) - g_{14}(t-t^{\prime})g_{23}(t-t^{\prime})] \\
&-\sum_{\alpha{\beta}\gamma{\delta}}\int{dt_{\alpha}}\int{dt_{\beta}}\int{dt_{\gamma}}\int{dt_{\delta}} 
g_{1{\alpha}}(t-t_{\alpha})g_{2{\beta}}(t-t_{\beta}) \\
&\times \bck{\alpha{\beta}}{\Gamma^{\text{4-pt}}(t_{\alpha},t_{\beta},t_{\gamma},t_{\delta})}{\gamma{\delta}} 
g_{\gamma{3}}(t_{\gamma}-t^{\prime})g_{\delta{4}}(t_{\delta}-t^{\prime}).
\end{split}
\end{equation}
The two-time reductions imply the propagation of either two particles (if $t>t^{\prime}$, giving an intermediate state $\Psi^{N+2}$), or two holes (if $t<t^{\prime}$, the intermediate state then being $\Psi^{N-2}$). For the moment we concentrate on the free part of the propagator, that is, the propagation of two non-interacting particles (as seen from the Eqs.~(\ref{eq:pqGamma4pt}), (\ref{eq:pqL4pt}) and (\ref{eq:pqR4pt}), only this part is needed to calculate the Parquet equations). We Fourier transform the first terms of the expression in Eq.~\ref{eq:Gpphh_4pt} to obtain the free ladder propagator $\mathcal{G}_{0}^{pphh}$. It can be found by inserting the inverse of the Fourier transform of the four-point Green's function (Eq.~(\ref{eq:FTg4pt})) into the expression for the Fourier transform and taking the appropriate limit. Since $\mathcal{G}_0^{pphh}$ is a function of one time difference only, we expect the Fourier transform to be a function of one energy only. As we will see below, the relevant total energy is the sum of the energy of state 1 and state 2, making coupling in the [12] channel the natural choice for this operator, namely
\begin{equation}\label{eq:Gpphh0}
\begin{split}
\bck{12}{\mathcal{G}_0^{pphh}(\Omega)}{34}=&\int{d(t_1-t_3)}e^{i\Omega(t_1-t_3)}\bck{12}{\mathcal{G}_0^{pphh}(t_1-t_3)}{34}=\\
=&\int{d(t_1-t_3)}e^{i\Omega(t_1-t_3)}\lim_{t_2\rightarrow{t_1}}\lim_{t_4\rightarrow{t_3}}\bck{12}{K_0(t_1,t_2,t_3,t_4)}{34} \\
=&\int{d(t_1-t_3)}e^{i\Omega(t_1-t_3)}e^{-i(\omega_1+\omega_2)t_1}e^{i(\omega_3+\omega_4)t_3}\int{\frac{d\omega_1}{2\pi{}}}\int{\frac{d\omega_2}{2\pi{}}}\int{\frac{d\omega_3}{2\pi{}}}\int{\frac{d\omega_4}{2\pi{}}} \\
&2\pi{}\delta(\omega_1+\omega_2-\omega_3-\omega_4)
i\bigl{[}2\pi{}\delta(\omega_1-\omega_3){g_{13}(\omega_1)}{g_{24}(\omega_2)}-2\pi{}\delta(\omega_1-\omega_4){g_{14}(\omega_1)}{g_{23}(\omega_2)}\bigr{]}.
\end{split}
\end{equation}
Changing variables in Eq.~(\ref{eq:Gpphh0}) to $\Omega=\omega_1+\omega_2, \omega=\omega_1-\frac{\omega_1+\omega_2}{2}, \omega^{\prime}=\omega_3-\frac{\omega_1+\omega_2}{2}$, we obtain
\begin{equation} \label{eq:Gpphh_energy_int}
\bck{12}{\mathcal{G}_{0}^{pphh}(\Omega)}{34}= 
i\int{\frac{d\omega}{2\pi{}}}[g_{13}(\Omega/2+\omega){g_{24}(\Omega/2-\omega)}-g_{14}(\Omega/2-\omega){g_{23}(\omega/2-\omega)}]. 
\end{equation}
To proceed, we insert the expression in Eq.~(\ref{eq:1p_prop_omega}) for the one-particle propagator. To ease readability, we use the abbreviation $z_{\alpha{\beta}}^{n+}$ for the overlaps $\bck{\Psi_0^{N}}{c_{\alpha}}{\Psi_n^{N+1}}\bck{\Psi_n^{N+1}}{c_{\beta}^{\dagger}}{\Psi_0^{N}}$ and so on. The integrals are simple contour integrals. Of the four terms, two have poles on the same side of the imaginary axis, and we close these on the opposite half plane so that they do not contribute to the integral, leaving two contributions evaluated by using the Residue theorem. Thus we obtain the following expression for the free ladder propagator $\mathcal{G}_0^{pphh}(\Omega)$:
\begin{multline} \label{eq:Gpphh0_energy}
\bck{12}{\mathcal{G}_{0}^{pphh}(\Omega)}{34}=i\int{\frac{d\omega}{2\pi{}}} 
\Biggl{[}\Bigl{[}\bigl{(}\sum_{n}\frac{z_{13}^{n+}}{\Omega/2+\omega-\epsilon_n^{+}+i\eta} + \sum_{k}\frac{z_{13}^{k-}}{\Omega/2+\omega-\epsilon_k^{-}-i\eta}\bigr{)} \\
\times \bigl{(}\sum_{m}\frac{z_{24}^{m+}}{\Omega/2-\omega-\epsilon_m^{+}+i\eta} + \sum_{l}\frac{z_{24}^{l-}}{\Omega/2-\omega-\epsilon_l^{-}-i\eta}\bigr{)}\Bigr{]}  \\
- \Bigl{[}\bigl{(}\sum_{n}\frac{z_{14}^{n+}}{\Omega/2+\omega-\epsilon_n^{+}+i\eta} + \sum_{k}\frac{z_{14}^{k-}}{\Omega/2+\omega-\epsilon_k^{-}-i\eta}\bigr{)}\\
\times \bigl{(}\sum_{m}\frac{z_{23}^{m+}}{\Omega/2-\omega-\epsilon_m^{+}+i\eta} + \sum_{l}\frac{z_{23}^{l-}}{\Omega/2-\omega-\epsilon_l^{-}-i\eta}\bigr{)} \Bigr{]} \Biggr{]} \\
 = \sum_{nm}\frac{z_{13}^{n+}z_{24}^{m+}}{\Omega-\epsilon_n^{+}-\epsilon_m^{+}+i\eta^{\prime}} - \sum_{kl}\frac{z_{13}^{k-}z_{24}^{l-}}{\Omega-\epsilon_k^{-}-\epsilon_l^{-}-i\eta^{\prime}} \\
+ \sum_{nm}\frac{z_{14}^{n+}z_{23}^{m+}}{\Omega-\epsilon_n^{+}-\epsilon_m^{+}+i\eta^{\prime}} - \sum_{kl}\frac{z_{14}^{k-}z_{23}^{l-}}{\Omega-\epsilon_k^{-}-\epsilon_l^{-}-i\eta^{\prime}}.
\end{multline}

\subsubsection{The ring propagator} 

We define the particle-hole ring propagator $\bck{12}{\mathcal{G}^{ph}(t-t^{\prime})}{34}$ as the reduction
\begin{multline} \label{eq:Gphdef}
\bck{12}{\mathcal{G}^{ph}(t-t^{\prime})}{34}{\equiv}\lim_{t_4\rightarrow{t}}\lim_{t_3\rightarrow{t^{\prime}}}\bigl{[}\bck{12}{K(t,t^{\prime};t_3,t_4)}{34}\\
 - \bck{\Psi^{N}_0}{c^{\dagger}_{4}(t_4)c_{1}(t)}{\Psi^{N}_0} \bck{\Psi^{N}_0}{c^{\dagger}_{3}(t_3)c_{2}(t^{\prime})}{\Psi^{N}_0}\bigr{]}\\
=\lim_{t_4\rightarrow{t}}\lim_{t_3\rightarrow{t^{\prime}}}\bigl{[}-i\bck{\Psi^{N}_0}{\mathcal{T}[c_{2_H}(t^{\prime})c_{1_H}(t)c^{\dagger}_{3_H}(t^{\prime})c^{\dagger}_{4_H}(t)]}{\Psi^{N}_0} \\
- \bck{\Psi^{N}_0}{c^{\dagger}_{4}(t)c_{1}(t)}{\Psi^{N}_0} \bck{\Psi^{N}_0}{c^{\dagger}_{3}(t^{\prime})c_{2}(t^{\prime})}{\Psi^{N}_0}\bigr{]}. 
\end{multline}
The incoming and outgoing states could in principle be any states, as the operator sequence in the propagator ensures that only state combinations of a hole-particle or particle-hole type give a non-zero expectation value.  
 
The last term in Eq.~(\ref{eq:Gphdef}) results from the fact that the exchange part of the propagator closes the diagrams into unconnected ground state energy diagrams, involving the one-body density matrix elements. We do not want to include these in the interaction operator, hence the definition (which in the literature often is called the polarization propagator). The proper exchanges of the interaction operator diagrams are generated automatically when antisymmetrized matrix elements are employed.

If we insert Eq.~(\ref{eq:4pointgreen+Gamma}) into the expression for $\mathcal{G}^{ph}$ in Eq.~(\ref{eq:Gphdef}), we obtain  $\mathcal{G}^{ph}$ in terms of the interaction operator $\Gamma^{\text{4-pt}}$:
\begin{multline} \label{eq:Gph_4pt}
\bck{12}{\mathcal{G}^{ph}(t-t^{\prime})}{34}=i[g_{13}(t-t^{\prime})g_{24}(t-t^{\prime}) 
-g_{14}(t-t^{\prime})g_{23}(t-t^{\prime})] \\
-\sum_{\alpha{\beta}\gamma{\delta}}\int{dt_{\alpha}}\int{dt_{\beta}}\int{dt_{\gamma}}\int{dt_{\delta}} 
g_{1{\alpha}}(t-t_{\alpha})g_{2{\beta}}(t-t_{\beta}) \\
\times \bck{\alpha{\beta}}{\Gamma^{\text{4-pt}}(t_{\alpha},t_{\beta},t_{\gamma},t_{\delta})}{\gamma{\delta}} 
g_{\gamma{3}}(t_{\gamma}-t^{\prime})g_{\delta{4}}(t_{\delta}-t^{\prime}).
\end{multline}
Looking at the free part of the propagator for the moment, we find the expression for $\mathcal{G}_{0}^{ph}$ as a function of energy in a similar manner as for the ladder propagator. This time we define $\Omega=\omega_1-\omega_3, \omega=\omega_1-\frac{\omega_1-\omega_3}{2}, \omega^{\prime}=\omega_2+\frac{\omega_1-\omega_3}{2}$. Then
\begin{equation} \label{eq:Gph0_energy_int}
\bck{12}{\mathcal{G}_{0}^{ph}(\Omega)}{34}=i{}\int{\frac{d\omega}{2\pi{}}}[g_{13}(\omega+\Omega/2){g_{24}(\omega-\Omega/2)}]. 
\end{equation}
We then insert the expression for the single particle propagators from Eq.~(\ref{eq:1p_prop_omega}) into the above expression, with the same notation for the overlaps as for the $\mathcal{G}^{pphh}$ calculation. Due to the sign of the energy variable, the two particle-hole terms survive in this case:
\begin{multline} \label{eq:Gph0_energy}
\bck{12}{\mathcal{G}_0^{ph}(\Omega)}{34}=-i\int{\frac{d\omega}{2\pi{}}}
\bigl{[}(\sum_{n}\frac{z_{13}^{n+}}{\omega+\Omega/2-\epsilon_n^{+}+i\eta} + \sum_{k}\frac{z_{13}^{k-}}{\omega+\Omega/2-\epsilon_k^{-}-i\eta})  \\
\times (\sum_{m}\frac{z_{24}^{m+}}{\omega-\Omega/2-\epsilon_m^{+}+i\eta} + \sum_{l}\frac{z_{24}^{l-}}{\omega-\Omega/2-\epsilon_l^{-}-i\eta}) \bigr{]} \\
 = \sum_{km}\frac{z_{13}^{k-}z_{24}^{m+}}{\Omega-\epsilon_k^{-}+\epsilon_m^{+}+i\eta^{\prime}} - \sum_{nl}\frac{z_{13}^{n+}z_{24}^{l-}}{\Omega+\epsilon_l^{-}-\epsilon_n^{+}+i\eta^{\prime}}.
\end{multline}
Note the difference in the definition of $\Omega$ between this expression and the expression for $\mathcal{G}^{pphh}(\Omega)$.

\subsection{Self-consistent Parquet equations} \label{subsec:selfparquet}

We are now ready to write down the Parquet equations in a formulation suitable for implementations, in our case a formulation which depends only on one energy. In doing this, we make the approximation that the total energy $\Omega$ is the same in the case of the ladder and ring propagators. The energy-dependence of the interaction can be seen as representing an average incoming energy.

Writing the equation for $R$ in the [13]-coupled notation given in Eq.~(\ref{eq:<13>}), we see that the Parquet equation can be rewritten in a compact matrix formulation as
\begin{eqnarray} \label{eq:parquet}
\bigl{[}\Gamma(\Omega)\bigr{]}&=& \bigl{[}I(\Omega)\bigr{]}+\bigl{[}L(\Omega)\bigr{]}+\bigl{[}R(\Omega)\bigr{]}, \nonumber \\
\bigl{[}L(\Omega)\bigr{]}&=& \bigl{[}(I+R)(\Omega)\bigr{]}\bigl{[}\mathcal{G}_0^{pphh}(\Omega)\bigr{]}\bigl{[}(I+R+L)(\Omega)\bigr{]}, \\
\bigl{[}R(\Omega)\bigr{]}&=& \bigl{[}(I+L)(\Omega)\bigr{]}\bigl{[}\mathcal{G}_0^{ph}(\Omega)\bigr{]}\bigl{[}(I+L+R)(\Omega)\bigr{]}.\nonumber
\end{eqnarray}
The equation for the self energy, given in Eq.~(\ref{eq:Sigma(E)}), connects the self energy with the interaction operator:
\begin{multline} \label{eq:par_Sigma(E)}
\Sigma(1,2;\omega)= -i \int_{C\uparrow}\frac{d\omega_1}{2\pi}\sum_{\alpha{\beta}}\bck{1{\alpha}}{V}{2{\beta}}g_{\alpha{\beta}}(\omega_1)\\
+ \frac{1}{2}\int{\frac{d\omega_1}{2\pi}}\int{\frac{d\omega_2}{2\pi}}\sum_{\alpha{\beta}\gamma{\delta}\mu{\nu}}\bck{1{\alpha}}{V}{\beta{\gamma}}g_{\beta{\delta}}(\omega_1)g_{\gamma{\mu}}(\omega_2) \\
\times \bck{\delta{\mu}}{\Gamma^{\text{4-pt}}(\omega_1,\omega_2,\omega,\omega_1+\omega_2-\omega)}{2{\nu}}g_{\nu{\alpha}}(\omega_1+\omega_2-\omega).
\end{multline}
When we approximate the full interaction operator $\Gamma^{\text{4-pt}}$ in this equation by the Parquet interaction operator $\Gamma = I+L+R$ given in Eq.~(\ref{eq:parquet}), we need to check whether we do any double-counting or not. When $I$ consists of only the bare interaction, Jackson {\it et al.}~\cite{Jackson1982} have shown that only contributions having the bare interaction $V$ as a top rung of a ladder term can be included. The other terms lead to double-counting either because they are simply equal to an already included term, or because the diagram is equal to an included term with some self-energy insertion, and therefore must be excluded. Thus the correct propagator is the $\mathcal{G}^{pphh}_0$ propagator and the correct coupling order is the [12] coupling, resulting in the following equation for the self energy:
\begin{multline}\label{eq:Sigmaparq}
\Sigma(1,2;\omega)= -i \int_{C\uparrow}\frac{d\omega^{\prime}}{2\pi}\sum_{\alpha{\beta}}\bck{1{\alpha}}{V}{2{\beta}}g_{\alpha,\beta}(\omega^{\prime})\\
+ \frac{1}{2}\int{\frac{d\omega^{\prime}}{2\pi}}\sum_{\alpha{\beta}\gamma{\delta}\mu{\nu}}\bck{1{\alpha}}{V}{\beta{\gamma}}\bck{\beta{\gamma}}{G^{pphh}_0(\omega+\omega^{\prime})}{\delta{\mu}}\bck{\delta{\mu}}{\Gamma(\omega+\omega^{\prime})}{2{\nu}}g_{\nu{\alpha}}(\omega^{\prime}).
\end{multline}
The single-particle propagator is expressed by the self energy via the Dyson equation, see Eq.~(\ref{eq:dyson}), repeated here for easy reference
\begin{equation} \label{eq:par_dyson}
g_{\alpha{\beta}}(\omega) = g^{0}_{\alpha{\beta}}(\omega)+\sum_{\gamma{\delta}}g^{0}_{\alpha{\gamma}}(\omega)\Sigma(\gamma,{\delta};\omega)g_{\delta{\beta}}(\omega).
\end{equation}
The propagators $\mathcal{G}_0^{pphh}$ and $\mathcal{G}_0^{ph}$ in the Parquet Eqs.~(\ref{eq:parquet}) are expressed by the amplitudes $z_{\alpha{\beta}}$ and excitation energies $\epsilon$ in the single-particle propagator found from solving the Dyson equation, creating complex dependencies which have to be solved iteratively.

\begin{figure}[ht]
\includegraphics[width=13cm]{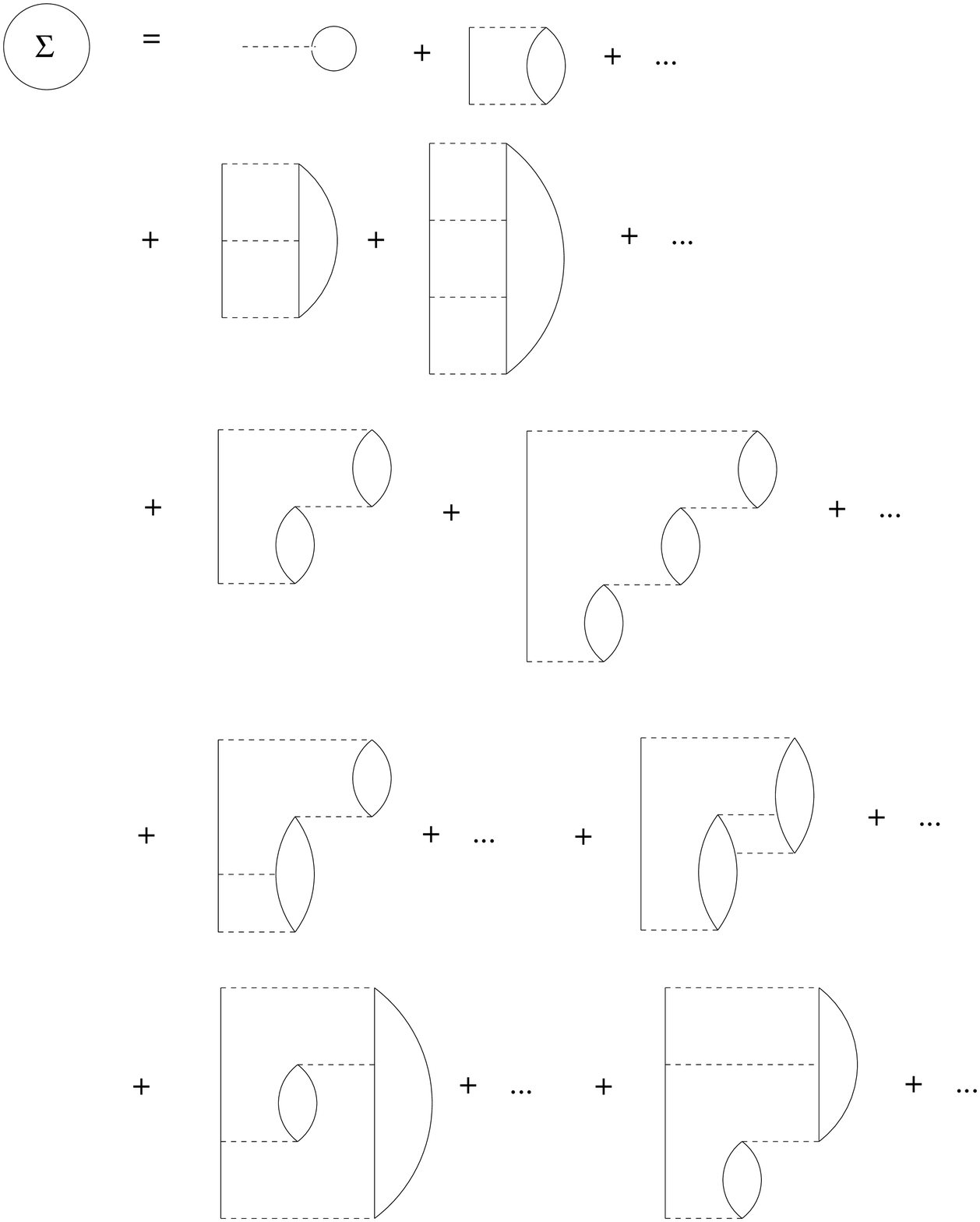}
\caption{The self energy diagrams generated by the Parquet method. All contributions to the fourth order are explicitly drawn. The propagators are dressed propagators.} \label{fig:pqsigma}
\end{figure}
\begin{figure}[ht]
\includegraphics[width=13cm]{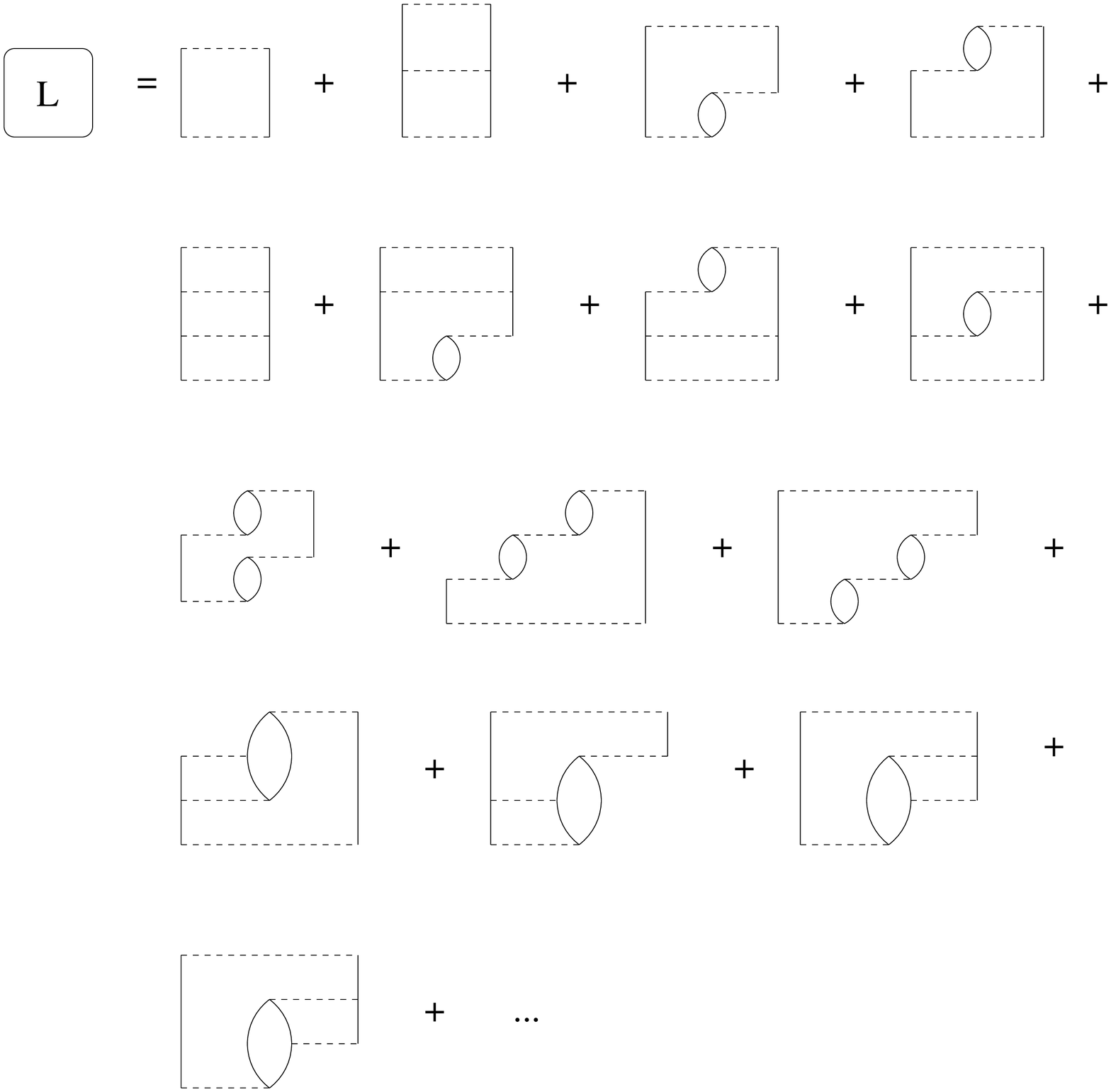}
\caption{The ladder diagrams generated by the Parquet method. All contributions to fourth order are explicitly drawn. The propagators are dressed propagators.} \label{fig:pqladder}
\end{figure}
\begin{figure}[ht]
\includegraphics[width=13cm]{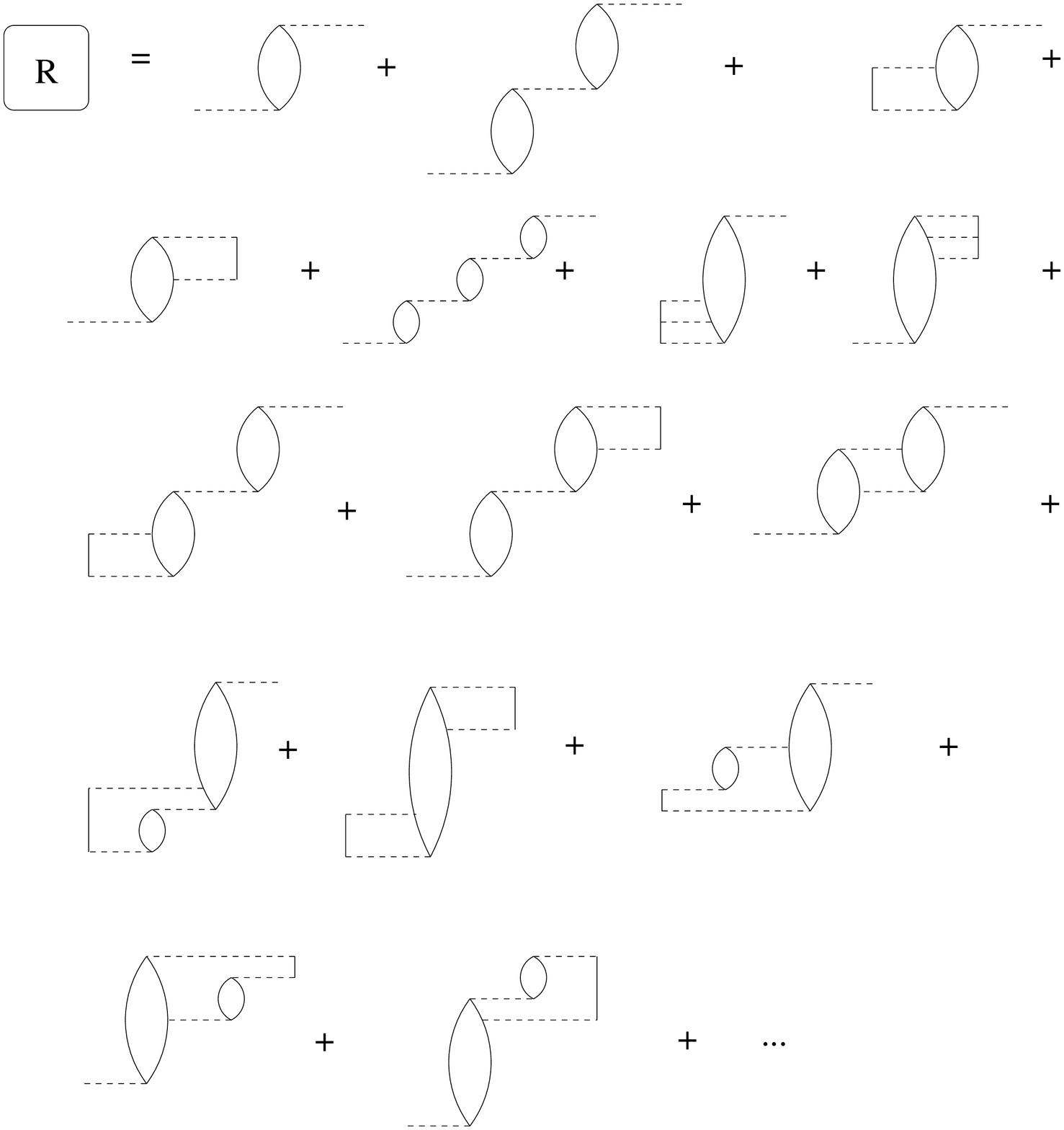}
\caption{The ring diagrams generated by the Parquet method. All contributions to fourth order are explicitly drawn. The propagators are dressed propagators.} \label{fig:pqring}
\end{figure}

Diagrammatically, all fourth order diagrams of the Parquet contributions to the self energy  are shown in figure.~\ref{fig:pqsigma}. All diagrams to fourth order for the ladder term are shown in Fig.~\ref{fig:pqladder}, and for the ring term in Fig.~\ref{fig:pqring}. The propagators are fully dressed propagators. 
Iteration by iteration, the diagrams are not generated order by order in the interaction, but rather staggered, all diagrams to fourth order being generated after three iterations.

We can compare the self energy diagrams to fourth order shown in Fig.~\ref{fig:pqsigma} with the corresponding Goldstone diagrams by closing the diagram by a hole line. Then we find that all Goldstone diagrams to fourth order are generated, in addition to several higher-order contributions as well (we remind the reader that the propagators are dressed, that is, all self energy insertions are included). Thus the Parquet method include ground state energy diagrams to the same level of precision as a coupled-cluster calculation including excitation operators to the fourth order, a CCSDTQ calculation~\cite{Bartlett1981}.

\section{Approximations in the numerical implementation}\label{sec:itersolution}

The equation set (\ref{eq:parquet}) and (\ref{eq:Sigmaparq}) together with the Dyson Eq.~(\ref{eq:dyson}) gives a solution to the one-body propagator if solved self-consistently. To find a viable implementation, we have to make some further simplifications, especially in the treatment of the energy variables.  

We have employed a number of approximations to the complete Parquet solution. The first is to truncate the infinite sum of diagrams in the class $I$ of diagrams which are non-simple in all channels after the first term, keeping only the bare interaction. The effect of missing the corrections stemming from the missing diagrams of fifth order or higher is small compared to the effects of some of our other approximations.

We have also made an approximation on the energy dependence of our interaction $\Gamma$, namely that the energy parameter $\Omega$ is the same in the calculation of $L$ and $R$, even though it is defined differently relative to the energies of the incoming states in the two cases. The $\Omega$ parameter thus represents an average incoming energy.
 
The small imaginary part $i{\eta}$ in the propagator Eqs.~(\ref{eq:Gpphh0_energy}) and (\ref{eq:Gph0_energy}) is a mathematical necessity to ensure that the correct poles of the propagator is included when integrating over $\omega$, and the real physics occurs in the limit of $\eta{\rightarrow}0$. Ideally, this should be done before a numerical implementation, but we have found that the poles of the propagator give rise to serious convergence problems, and so the $i{\eta}$ factor must be present also in the numerical implementation. The physical limit is then found as an extrapolation of results for different values of $\eta$. The effect of $\eta$ is to move the poles away from the real axis. This reduces the severity of the poles and smoothens the energy dependence of the real matrix elements in exchange for larger imaginary parts. Thus the overall effect is to reduce the effects of the interaction between the particles, making the results closer to a mean field result. This need for an extrapolation introduces an additional uncertainty to our results.

Our method for handling the poles and the Dyson equation is different from the implementation of the Green's functions approach employed by Barbieri {\it et.al.}~\cite{Barbieri2004,Barbieri2006,Barbieri2009}.  They solve the Dyson equation numerically, and then use a small number of solutions close to the Fermi level, typically two or three. The remaining solutions are accounted for in an average manner. This simplifies the calculation considerably. In their approach, the interaction $\Gamma$ is handled in the Faddeev random phase approximation.

The most influential approximation is our approximate method for solving the Dyson equation. As discussed  above, in the exact solution, the standard Hartree-Fock quasi-particles and quasi-holes are no longer stable single-particle states. The energy of a single-particle state gets smeared out over a broad range of possible energies. While it is possible to find such multiple solution sets, see for example Ref.~\cite{Barbieri2002a, Barbieri2001,Barbieri2002,Barbieri2003}, the ensuing complexity makes computations within our scheme far too demanding at present. We have therefore opted for an approximation where we keep only the solution closest to the first order energy $\epsilon^{f.o.}$. The first order energies are determined from the energy-independent first order contribution to $\Sigma$
\be  \label{eq:f.o.Sigma}
\Sigma^{f.o.}(\alpha,\beta)= -i \int_{C\uparrow}\frac{d\omega}{2\pi}\sum_{\gamma{\delta}}\bck{\alpha{\gamma}}{V}{\beta{\delta}}g_{\gamma,\delta}(\omega),
\ee
and
\be \label{eq:epsilon_f.o.}
\epsilon^{f.o.}_{\alpha{\beta}}=e^0_{\alpha{\beta}}+\Sigma^{f.o.}(\alpha{\beta}).
\ee
The energy of a given single-particle state is calculated from Eq.~(\ref{eq:epsilon(omega)}) with $\Sigma$ calculated at the first order energy, that is, the energies are the eigenvalues of the equation
\be \label{eq:Dysonapprox}
([e]+[\Sigma(\epsilon^{f.o.})])\ket{\lambda}=\omega_{\lambda}\ket{\lambda},
\ee
which is now a simple, linear eigenvalue problem. The number of eigenvalues is then limited to $N$, the number of orbitals in the basis. As a consequence, the hole spectral function in Eq.~(\ref{eq:Sh}), will be given by
\begin{equation}\label{eq:Sh_pq}
S_{h}(\alpha,\omega) = \sum_{k} |z_{\alpha{\alpha}}^{k-}|^{2}\delta(\omega-\epsilon_k^{-}).
\end{equation}
The sum over $k$ in this equation is limited to the number of orbitals, and the energies $\epsilon_k^{-}$ are the orbital energies. The spectroscopic factors will be smaller than 1, as the coupling between states with different orbital numbers $n$ will give hole spectral functions which have some probability of having a higher energy. As each energy can be identified with a definite orbital, the height of the spike at that energy gives the spectroscopic factor of that orbital. The sum rule that the occupation and depletion numbers of a single basis state must sum to 1 reduces to the condition that the sum of the hole amplitude and particle amplitude for a given energy sums to 1, giving exactly complementary hole and particle spectral functions.

The integral over the energy of the single-particle propagator in Eq.~(\ref{eq:Sigmaparq}) is solved by the Residue theorem. In the first order term the interaction is energy-independent, and so no complications occur. To perform the integration in the second term, we make the assumption that the residues at the poles of the ladder propagator $\mathcal{G}_0^{pphh}$ and the interaction $\Gamma$ are small compared to the residue at the pole of the single-particle propagator. These residues will quickly be quenched by the denominator of the single-particle propagator as long as they are at other energies than the energies of the single-particle orbitals. Due to the approximate solution of the Dyson equation, the sums over $n+$ and $k-$ in the expression for the single particle propagator reduce to restricting the summations over orbitals to either over or under the Fermi level. Thus our expression for the self energy becomes
\begin{equation}\label{eq:Sigmaparq_app}
\begin{split}
\Sigma(1,2;\omega)= &  \sum_{\alpha{\beta}}\bck{1{\alpha}}{V}{2{\beta}}\sum_kz^{k-}_{\alpha\beta}\\
& + \frac{1}{2}\sum_{n+}\sum_{\alpha{\beta}\gamma{\delta}\mu{\nu}}\bck{1{\alpha}}{V}{\beta{\gamma}}\bck{\beta{\gamma}}{G^{pphh}_0(\omega+\epsilon_{\nu{\alpha}}^{n+})}{\delta{\mu}}\bck{\delta{\mu}}{\Gamma(\omega+\epsilon_{\nu{\alpha}}^{n+})}{2{\nu}}z_{\nu{\alpha}}^{n+} \\
& +\frac{1}{2}\sum_{k-}\sum_{\alpha{\beta}\gamma{\delta}\mu{\nu}}\bck{1{\alpha}}{V}{\beta{\gamma}}\bck{\beta{\gamma}}{G^{pphh}_0(\omega+\epsilon_{\nu{\alpha}}^{k-})}{\delta{\mu}}\bck{\delta{\mu}}{\Gamma(\omega+\epsilon_{\nu{\alpha}}^{k-})}{2{\nu}}z_{\nu{\alpha}}^{k-} \\
= &  \sum_{\alpha{\beta}<F}\bck{1{\alpha}}{V}{2{\beta}}z_{\alpha\beta}\\
& + \frac{1}{2}\sum_{\alpha{\nu}>F}\sum_{{\beta}\gamma{\delta}\mu}\bck{1{\alpha}}{V}{\beta{\gamma}}\bck{\beta{\gamma}}{G^{pphh}_0(\omega+\epsilon_{\nu{\alpha}}^{f.o.})}{\delta{\mu}}\bck{\delta{\mu}}{\Gamma(\omega+\epsilon_{\nu{\alpha}}^{f.o.})}{2{\nu}}z_{\nu{\alpha}} \\
& + \frac{1}{2}\sum_{\alpha{\nu}<F}\sum_{{\beta}\gamma{\delta}\mu}\bck{1{\alpha}}{V}{\beta{\gamma}}\bck{\beta{\gamma}}{G^{pphh}_0(\omega+\epsilon_{\nu{\alpha}}^{f.o.})}{\delta{\mu}}\bck{\delta{\mu}}{\Gamma(\omega+\epsilon_{\nu{\alpha}}^{f.o.})}{2{\nu}}z_{\nu{\alpha}}.
\end{split}
\end{equation}
Here $z_{\nu{\alpha}}$ is the amplitude of the single-particle propagator at the first-order energy $\epsilon_{\nu{\alpha}}^{f.o.}$. We have chosen this solution method to incorporate the effect of the changes in the spectral function into the self energy while still conserving the total number of particles.

The implication of the fixed-energy Dyson equation approximation is that while we loosen the restriction that the input basis states are 'good' states and allow the single-particles to become linear combinations of the chosen basis set, we still assume that the single-particle picture is valid, that is, the system can be described as a set of (quasi-)particles with a discrete energy spectrum.



We present results for two different schemes for handling the energy dependence of the Parquet equations. In the energy-independent scheme, a fixed starting energy $E_{in}$ is chosen and used as the input $\Omega$ in the equations for the two-time propagators. Typically, $E_{in}$ has to have a value well removed from the poles of the propagators to obtain converged results. The generated interaction $\Gamma(E_{in})$ is analogous to the interaction from a conventional $G$-matrix calculation, with some important differences. The ladder terms includes both particle-particle and hole-hole ladders, whereas the $G$-matrix only contains the particle-particle part. The starting energy could correspond to the energy of the incoming particles or to the energy difference between a particle-hole pair, depending on context. Like a conventional $G$-matrix, eventual use of the generated interaction would be hampered by the need to extrapolate for starting energies for which the poles of the propagator destabilize the solution (i.e. where the incoming particle energies correspond to a pole in the propagator). The inclusion of hole-hole terms imply that also negative starting energies will give this effect. In our approach this destabilization can be handled by increasing the $i{\eta}$ parameter significantly, thus generating an interaction applicable to all energies. 

In the energy-dependent scheme, a (real) energy grid is set up and the propagators and interactions are calculated at each value of the grid. The values of $\Gamma(\Omega)$ used in the calculation of the self energy (Eq.~(\ref{eq:Sigmaparq_app})) are then interpolated to the energy $\Omega = \omega + \epsilon_{\nu{\alpha}}^{f.o.}$. This scheme includes the poles in the generated interaction in a more correct manner than the energy-independent scheme, but the added number of poles gives additional convergence problems. 

\section{Application to a simple model}\label{sec:simple}

In this work we have chosen to test the formalism on a simple model which captures several basic features
of nuclei, such as a discrete single-particle spectrum, strong pairing correlations and core-polarization.
Applications to given nuclei will be presented in forthcoming articles. The aim here is to expose the formalism. 
The structures seen in more realistic cases are here much simpler to analyze, making it possible to discern the origins of the observed features of the self energy and the spectral functions to a greater extent, including the relative effects of the pair-correlation and the particle-hole terms in the interaction. Thus the insights gained in this work  are of relevance to the discussion of the more realistic cases as well. 

 We describe the model as built around a three-term Hamiltonian. In Section~\ref{subsec:pair_f_eta} we discuss the stability of the Parquet solutions with respect to the parameter $\eta$. In Section~\ref{subsec:selfres} we discuss the self energy, and the results for the spectral function is presented in Section~\ref{subsec:specfcnres}. We compare our results with an exact diagonalization in Section~\ref{subsec:compres}.

\subsection{Description of the model}\label{sec:pairmodel}

\begin{figure}[hbtp]
\includegraphics[width=12cm]{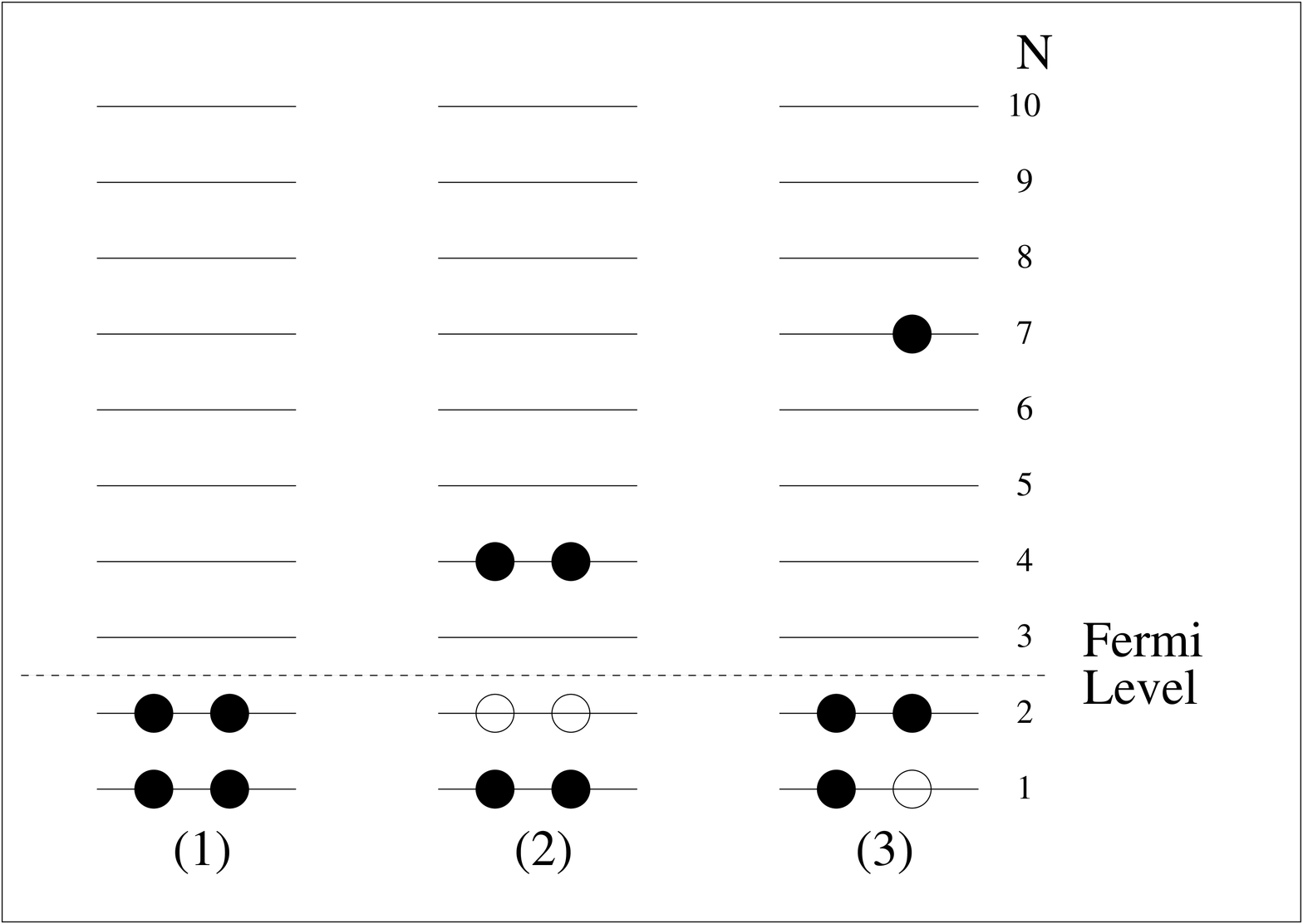}
\caption{Sketch of the $N=10$, $p=4$ model in the ground state~(1), an example pair excitation, the only type to occur when  $g\neq{0}$, $f=0$~(2), and an example pair-breaking excitation which can occur when $f\neq{0}$~(3).} \label{fig:pairmodel}
\end{figure}

The model has $N$ doubly-degenerate and equally spaced single-particle levels labelled by $n=0,\dots,N_{max}$ and spin $\sigma=\pm{1}$. The Fermi level defines the boundary of the ``closed core'', as shown in the first column of Fig.~\ref{fig:pairmodel}. We define a Hamiltonian of the system with three contributions, a one-body part $H_0$ and a two-body interaction $V$ consisting of two terms, as follows:
\begin{equation}
H = H_0 + V =
\sum_{k\sigma}kc_{k\sigma}^{\dagger}c_{k\sigma} + \frac{1}{2}g \sum_{kj}c_{k+}^{\dagger}c_{k-}^{\dagger}c_{j-}c_{j+} + \frac{1}{2}f \sum_{jkl}(c_{k+}^{\dagger}c_{k-}^{\dagger}c_{j-}c_{l+}+ c_{j+}^{\dagger}c_{l-}^{\dagger}c_{k-}c_{k+})
\end{equation}
The energy of the first level is set to 0, and the energy increases by a fixed amount for each level. We set this fixed level spacing to 1, and the coupling constants $g$ and $f$ give the relative strengths between the level spacing and the interaction. The first term of the interaction has a pair structure, and can only excite two particles at a time, as shown in the second column of Fig.~\ref{fig:pairmodel}. The vacant positions are holes and are drawn as open circles. The second term in the interaction is a pair-breaking term, as it acts between pairs of opposite spins, creating excitations of the type shown in the third column of Fig.~\ref{fig:pairmodel}. 
This system can be solved exactly by diagonalization, enabling us to study the accuracy of the Parquet summation method. We pay particular  attention to the effects of the relative strength between the level spacing and the interaction, and to the effects of increasing the number of single-particle orbitals and particles. As our implementation of the Parquet method so far only includes two-body interactions, we can gain insights into the influence of many-body correlations beyond the two-body level. In nuclei, the pairing component of the interaction is known to be strong, as seen by the success of the seniority scheme models~\cite{Talmi1993}, and thus the pairing-only model is of interest in this context. We know that closed-core nuclei commonly have an interaction strength of $\sim{20-30}$\% of the level spacing \cite{Dean2003}, so we will concentrate on the span $[-1,1]$ of interaction strengths. 

We employ dimensionless variables in the discussion of this model.

\subsection{Convergence with respect to $\eta$}\label{subsec:pair_f_eta}

Since our method is based on an iterative procedure, we need to investigate the stability of the solution and see if the results converge as the number of iterations increase. The $\eta$ parameter in the two-time propagators~(\ref{eq:Gpphh0_energy}) and (\ref{eq:Gph0_energy}) regulates the influence of the pole terms, determining the stability of the iterative procedure. Increasing values of $\eta$ give calculations in which the effect of the poles are increasingly removed, increasing the imaginary parts of the results.

\begin{figure}[hbtp]
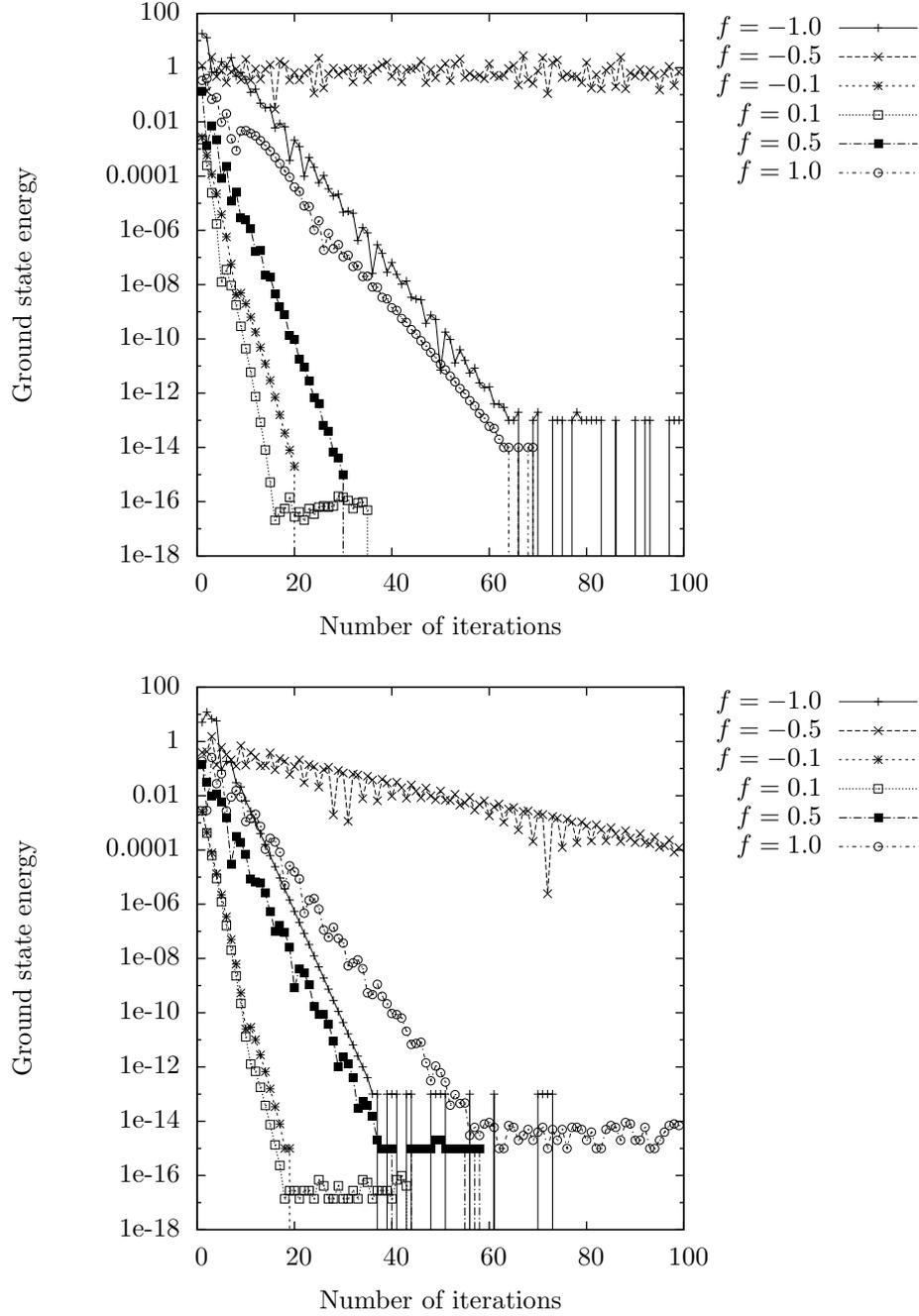

\begin{tabular}{c}
\input{fhfn10p2ed1et1.tex} \\
\input{fhfn10p2ed1et5.tex} 
\end{tabular}
\caption{The difference $E_n-E_{n-1}$ between successive iterations as a function of the number of iterations for energy-independent calculations with $E_{in}=-20$ for several $f$ values. with $\eta=1$ (upper panel) and $\eta=5$ (lower panel) in the $N=10$, $p=2$ model.} 
\label{fig:p_fhf_n10p2}
\end{figure}

For the simplest case, a two-level pairing-only model ($f=0$, $N=2$ and $p=2$), the self energy becomes diagonal. The unperturbed single-particle propagator structure as given in Eq.~(\ref{eq:1p_prop_free}) is conserved. The parameter $\eta$ can be set to 0 in most cases for energy-independent calculations within this simplest model. Most starting energies give convergence to machine precision within 10-15 iterations. The exception to this is starting energies exactly at the poles in the $\mathcal{G}_0^{pphh}$ and $\mathcal{G}_0^{ph}$ propagators used in the calculation of $L$ and $R$, respectively. The convergence properties for the $\eta=0$ calculations are less stable when the pairing constant $|g|$ increases. Energy-dependent calculations have roughly the same convergence properties as the energy-independent case. Except for some unhappy cases where an exact pole in $\Gamma$ is encountered, the results for all $g$ values converge for $\eta=0$. Setting $\eta>0$ gives faster convergence.

Increasing the system size changes the convergence properties, as the number of poles increases. For the $N=10$, $p=4$ system energy-independent calculations converge to machine precision within 25 iterations for a starting energy of $E_{in}=-20$. The energy-dependent calculations exhibit an even slower convergence pattern compared with the $N=2$ case.
 
Setting $f\neq{0}$ changes the interaction from a purely pair-conserving to a pair-breaking interaction. For simplicity, we set $g=0$ when discussing the impact of pair-breaking. This amounts to having a pair-breaking contribution equal in size to the pair-conserving contribution. The pair-breaking term introduces some new, interesting features to the model, most notably in the fact that the self-energy is no longer diagonal.

For the energy-independent scheme, no convergent solutions can be found in the $N=2$, $p=2$ system for values of $f>0$ when $\eta=0$. This is due to the closeness of the first-order energies of the two levels. At the value $f=1$ they become equal. Setting $\eta>0$ remedies the instability for values of $f\lesssim{0.6}$. Above $f\sim{0.6}$ the solutions start to oscillate as a function of the number of iterations for all $\eta$.

In the energy-dependent scheme, convergence is good for energy mesh grids with 10-30 points. Increasing the number of points further destabilizes the solution for some values of $f$. Then some of the mesh points hits the poles in $\Gamma$ due to the two-time propagators. The $\eta=0$ calculations do not converge. Increasing $\eta$ remedies this, then an almost exact match between all grid sizes above 10 is observed. The convergence is good for $\eta>0$ when $f<0.5$, as in the energy-independent case. For larger $f$ values, the instabilities leads to in general poor convergence properties in the energy-dependent scheme.

Increasing the number of levels to 10, negative values of $f$ (attractive interaction) become unstable in the range values of $f\sim{-0.4}$ to $f\sim{-0.6}$. An $\eta$ value as large as 5 is needed before convergence is achieved. In Fig.~\ref{fig:p_fhf_n10p2} we show the difference $E_n-E_{n-1}$ in a log-scale plot as a function of number of iterations. The calculations are done in the energy-independent scheme in the $N=10$, $p=2$ system for $\eta=1$ (upper panel) and $\eta=5$ (lower panel). The instability seen in the $N=2$ system at $f>0$ is not present any longer, that is, with higher $\eta$ values we see convergent results for all positive $f$ values up to 1. 

We observe that adding the imaginary component gives a more irregular convergence pattern. The graphs shown are representative for the general pattern in all the calculations.

The $N=10$, $p=4$ shows similar, but slightly less convergent patterns for positive $f$ values, and there is no instability around $f\sim{-0.5}$. All negative $f$ values are convergent at $\eta=1$.

The energy-dependent calculations are more unstable. For the  $N=10$, $p=2$ system, convergence could not be obtained for any value of $\eta$ when $f\leq{-0.7}$, and for the $N=10$, $p=4$ system, the limit for obtaining convergent results is $f\leq{-0.5}$. 

Generally, the number of particles seems to have a larger impact on the convergence properties when the pair-breaking term is included.

\subsection{Self energy}\label{subsec:selfres}

The pair model is ideal for studying the pole structure of the self energy, as it is easy to discern the effects of changing different parameters. The energy dependence of the self energy $\Sigma$ can be investigated both for the energy independent and energy dependent cases. The first order term in Eq.~(\ref{eq:Sigmaparq_app}) is energy-independent, giving the energies at which the self energy is calculated at each iteration. The energy-dependent term has poles stemming both from poles in the $\mathcal{G}_0^{pphh}$ propagator and from the interaction $\Gamma$, giving a rich structure in an exact calculation. The pair-conserving interaction gives a diagonal self energy.

\begin{figure}[hbtp]
\input{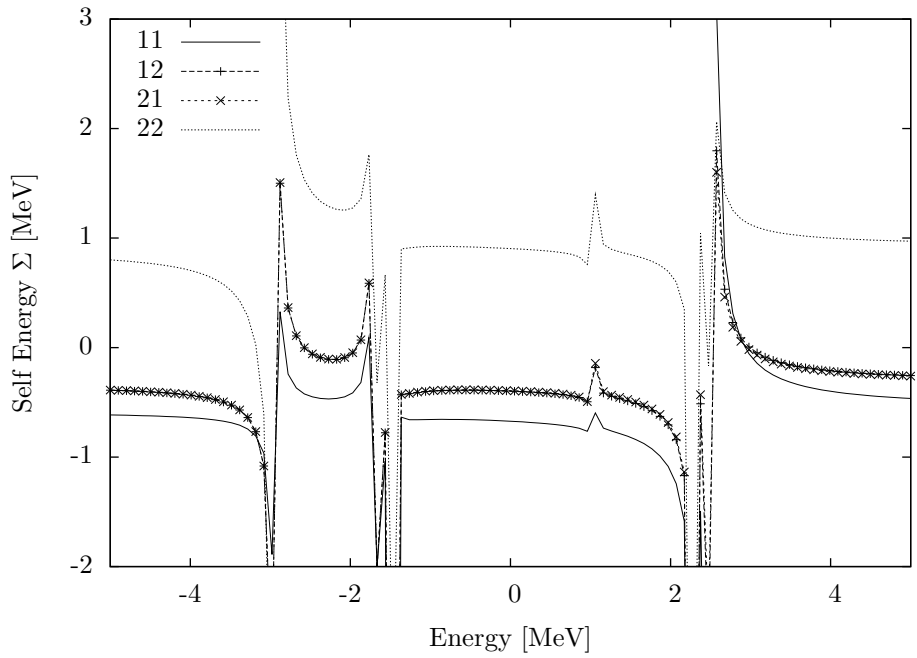}
\caption{Energy-independent calculation of ${\Sigma}$ as a function of energy calculated for $f=-0.5$ in the $N=2$, $p=2$ system. The legends shows the indices of $\Sigma$, that is,$\bck{1}{\Sigma}{1}$ and so forth, according to the levels in Fig.~\ref{fig:pairmodel}. } 
\label{fig:pSf_hf}
\end{figure}

The pair-breaking part of the interaction generates off-diagonal contributions to the first-order self energy, that is, $\bck{1}{\Sigma}{2},\bck{2}{\Sigma}{1}\neq{0}$ in the $N=2$, $p=2$ model. This is illustrated in Fig.~\ref{fig:pSf_hf} showing the self energy matrix elements obtained by an energy-independent calculation with $L=R=0$ and $f=-0.5$. The numbering of the states in the matrix elements is given by  the level numbers shown in Fig.~\ref{fig:pairmodel}. 

The $\mathcal{G}^{pphh}$ propagator has eight poles (the Dyson equation solution gives only two energies in each sum in Eq.~(\ref{eq:1p_prop_omega})), but some will probably be quenched by the small removal/addition amplitudes. Inclusion of an energy-independent $\Gamma$ will give pole positions which depend on the starting energy, as the poles in the two-time propagators also have this dependence. The position of the poles in the self energy is further adjusted by the self-consistency procedure, but due to our approximate solution to the Dyson equation, the number of poles remain unchanged.

Reducing the absolute value of the interaction strength parameter $f$ reduces the impact of the poles drastically. The main effect of an attractive force is to lower the average energy of the lowest-lying level and increasing the gap between the two levels. This lowers the ground state energy. The repulsive force reduces the gap by pushing the lowest level upwards in energy.

In the upper panel of Fig.~\ref{fig:pSf_n10p4} we have shown the matrix element $\bck{1}{\Sigma}{1}$ in the $N=10$, $p=4$ for calculations with $\eta=0$ and with $\eta=1$. Increasing $N$ and $p$, the number of poles increases as expected. We know that it is necessary to have $\eta>0$ to obtain convergence in the larger models. As the number of poles in the larger systems quickly becomes quite large, the effect of $\eta$ is to dampen the pole structure of the self energy to resemble an average self energy.  When $\eta$ becomes large, all structure is lost, and the solution found is a mean-field solution.  The solution is different from the first-order solution, however, as the self energies are complex, and there is a larger number of non-zero off-diagonal matrix elements. A convergence failure for all values of $\eta$ indicates that no such solution can be found within our scheme. 
\begin{figure}[hbtp]
\begin{tabular}{c}
\input{Sf_hf_n10p4_eta.tex} \\
\input{Sf_hf_n10p4.tex}
\end{tabular}
\caption{The upper panel shows the $\bck{1}{\Sigma}{1}$ matrix element in the $N=10$, $p=4$ system at $f=-0.5$ for $\eta=0$ and for $\eta=1$ after the first iteration. The lower panel shows the diagonal matrix elements $\bck{1}{\Sigma}{1}$ (labelled 1) and $\bck{3}{\Sigma}{3}$ (labelled 3), and the off-diagonal elements $\bck{1}{\Sigma}{2}$ (labelled 12) and $\bck{2}{\Sigma}{1}$ (labelled 21). The off-diagonal elements are not equal.} 
\label{fig:pSf_n10p4}
\end{figure}

The self energy matrix elements show a characteristic separation between the two groups of poles. The pole structure for positive energies is due to two-particle-one-hole excitations, while the pole structure at negative energies is due to two-hole-one-particle excitations.

We have shown diagonal matrix elements $\bck{1}{\Sigma}{1}$ and $\bck{3}{\Sigma}{3}$ of the $\eta=1$ calculation in the lower panel of Fig.~\ref{fig:pSf_n10p4}. These are representative for the self energy corresponding to holes ($\bck{1}{\Sigma}{1}$) and particles ($\bck{3}{\Sigma}{3}$). The off-diagonal element $\bck{1}{\Sigma}{2}$ and  $\bck{2}{\Sigma}{1}$ are representative for all the off-diagonal elements. They have values slightly below 0, and the remnants of pole structure is most prominent at negative energies. The average energies of the diagonal elements are determined by the input energy level. The structure at negative energies shifts to the left with increasing level number. Only the diagonal elements corresponding to hole states have a distinct shape stemming from two-particle-one-hole propagation. The pole structures of the hole diagonal elements and particle diagonal elements for the $\eta=0$ calculation have different average features (although both types show the characteristic two groups of poles seen in the $\eta=0$ graph of Fig.~\ref{fig:pSf_n10p4}), and this is reflected in the different shapes of the $\eta=1$ graphs.

An interesting feature to notice in the lower panel of Fig.~\ref{fig:pSf_n10p4} is the fact that the off-diagonal matrix elements are unequal, due to the off-diagonal elements of the single-particle propagator. Thus the self energy matrix in the Dyson equation is a non-symmetric matrix when the number of levels and particles have increased.
\begin{figure}[hbtp]
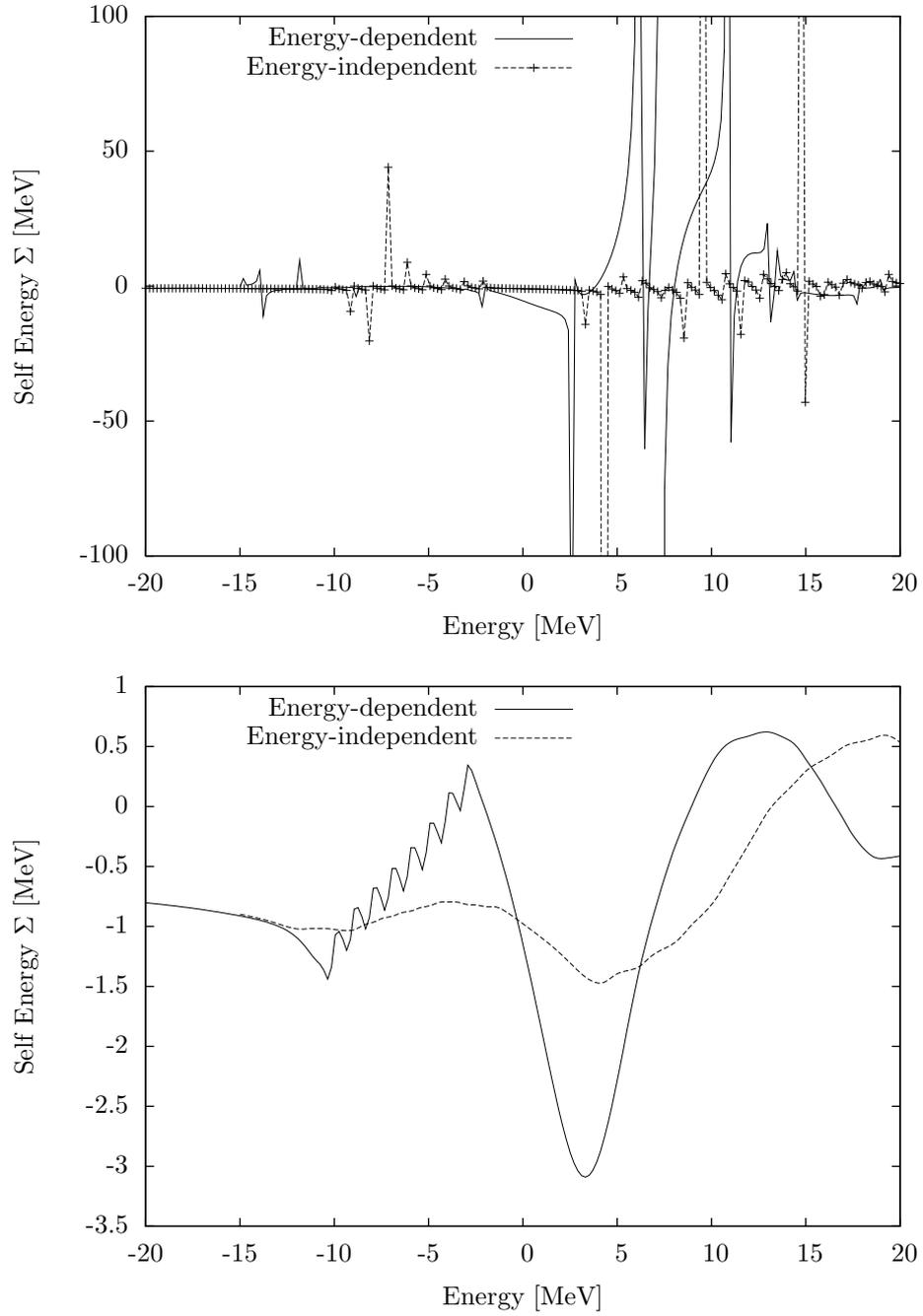

\begin{tabular}{c}
\input{Sf_up_n10p4_edim.tex}\\
\input{Sf_up_n10p4_edim_et1.tex}
\end{tabular}
\caption{The difference between the matrix element $\bck{1}{\Sigma}{1}$ of the self energy of an energy-independent and an energy-dependent calculation after 1 iteration in the $N=10$, $p=4$ system with $\eta=0$ in the upper panel and with $\eta=1$ in the lower panel.} 
\label{fig:pSfup_n10p4_edim}
\end{figure}

The $\bck{1}{\Sigma}{1}$ element in an energy-independent and an energy-dependent calculation with $\eta=0$ is given in the upper panel of Fig.~\ref{fig:pSfup_n10p4_edim}. The influence of the poles of the $L$ and $R$ contributions to $\Gamma$ makes the energy-dependent self energy more irregular and the poles more prominent for positive energies than in the energy-independent case. We observe that the poles at negative energies are shifted to the left. In the lower panel we have shown the same two matrix elements after the first iteration when $\eta=1$ in both calculations. The simpler structure and generally smaller amplitudes in the energy-independent $\eta=0$ case give a faster damping with $\eta$ than in the energy-dependent case, at least partly explaining the better convergence properties. The main contribution stemming from two-particle-one-hole propagation is shifted towards higher energies.

\subsection{Spectral functions}\label{subsec:specfcnres}

The single particle propagator has non-diagonal amplitudes, resulting in a discrete spectral function. The coupling between states with different orbital numbers will give hole spectral functions which have some probability of having an higher energy. As each energy can be identified with a definite orbital, the height of the spike at that energy gives the spectroscopic factor of that orbital. 
\begin{figure}[hbtp]
\input{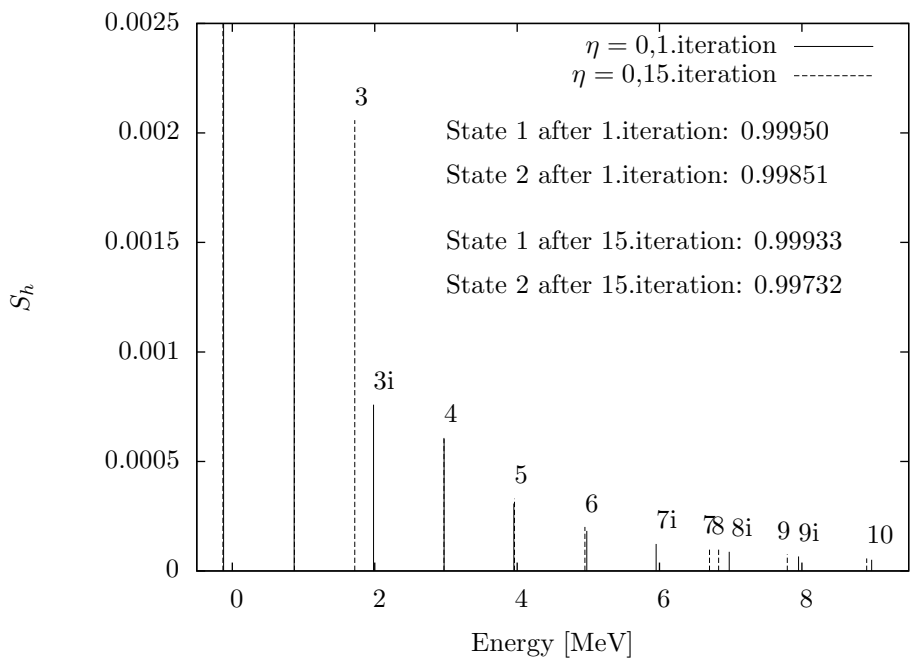}
\caption{Spectral function in the $N=10$, $p=4$ system with $f=-0.1$, after 1 and after 15 iterations. The numbers above each bar is the level associated with that energy (due to the approximation of the Dyson equation, such a correspondence can be made), according to the numbers of Fig.~\ref{fig:pairmodel}. The duplicate numbers 3i and 3 and so forth indicate the spectral function after 1 iteration and after 15 iterations, respectively. Only a small amount of the hole strength is distributed to the particle states at this small $|f|$ value.} 
\label{fig:pSpg_f-01}
\end{figure}
When the interaction is 0, the particles are independent, and the spectral function will have one spike of height 1 at the energy of each basis state. The hole spectral function has thus 1 at the levels chosen to be holes, and 0 at the levels chosen to be particles. The effect of changing the interaction strength $f$ is to reduce the height of the most dominant spikes, giving small amplitudes also at the energies of the other basis states, as shown in Fig.~\ref{fig:pSpg_f-01} for a $N=10$, $p=4$ system with $f=-0.1$. The state numbers are according to the level scheme in Fig.~\ref{fig:pairmodel} Only a small amount of the hole strength is distributed to the particle states at this small $|f|$ value. In the same figure we have also shown the hole spectral function obtained after 15 iterations, confirming that the process of self-consistency does not influence much on this function when the interaction strength is small. The particle spectral function exactly mirrors the hole function.

\begin{figure}[hbtp]
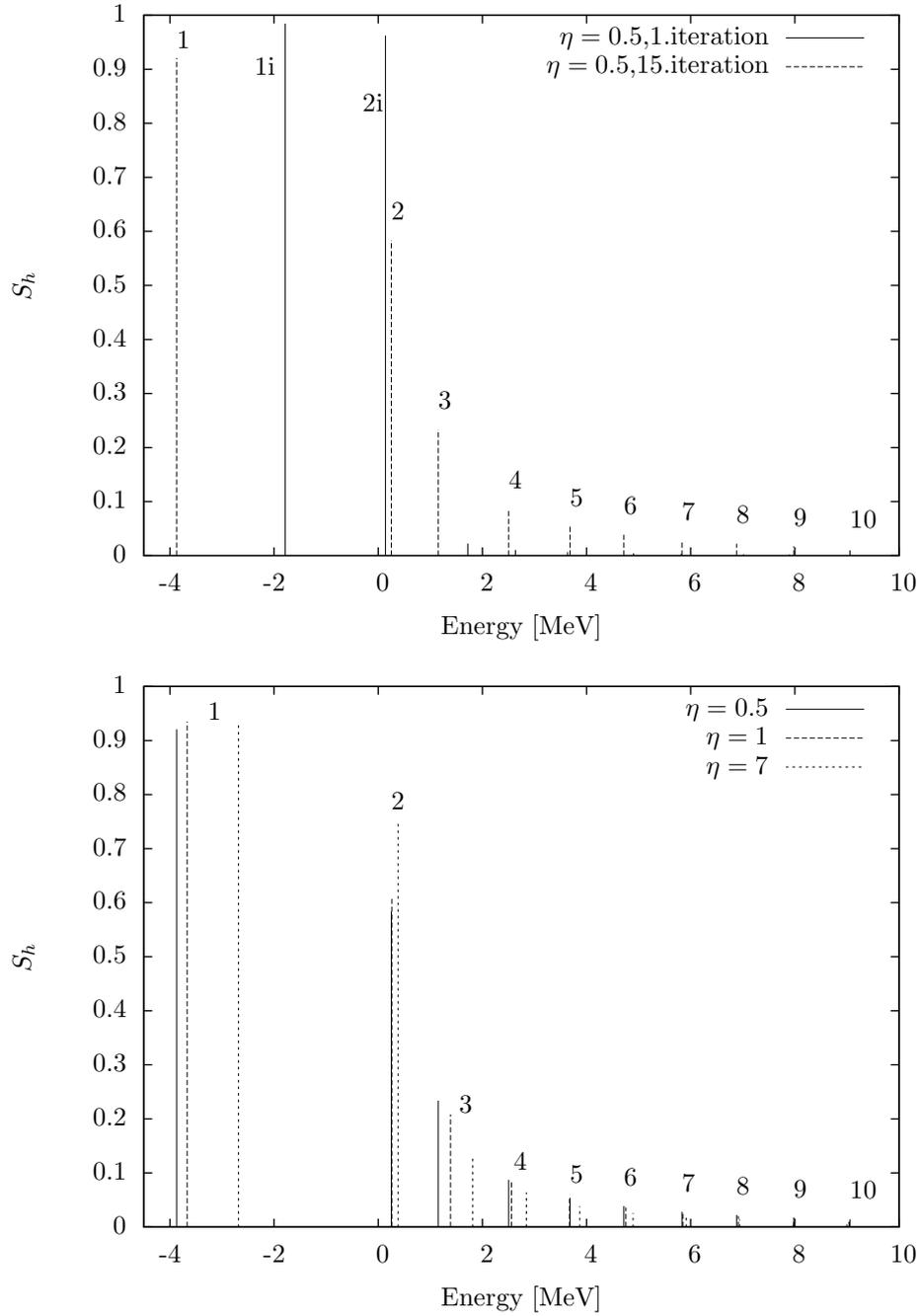

\flushleft
\begin{tabular}{c}
\input{Spg_p4f-05et05.tex} \\ 
\input{Spg_p4f-05eta.tex} \\ 
\end{tabular}
\caption{Spectral function in the $N=10$, $p=4$ system with $f=-0.5$. In the upper panel the differences between 1 and after 15 iterations is shown. The lower panel shows several $\eta$ values after 15 iterations. The state numbers are according to the level scheme in Fig.~\ref{fig:pairmodel}. The duplicate numbers 1i and 1 and so forth indicate the spectral function after 1 iteration and after 15 iterations, respectively. } 
\label{fig:pSpg_f-05et1}
\end{figure}

Increasing the interaction strength to $f=-0.5$ gives unstable results for $\eta=0$. In the upper panel of Fig.~\ref{fig:pSpg_f-05et1} we compare the results for the hole spectral function in the  $N=10$, $p=4$ system at $\eta=0.5$ after the first iteration and after 15 iterations. The two lowest-energy (hole) states starts out having a height close to 1, and there is very little strength on any of the higher-lying states. During the iteration procedure, the energy of the lowest state is considerably lowered, but the strength remains almost the same. The second hole state remains at almost the same energy, but the strength is reduced by around 30\%, and the strength of the higher states increased accordingly. Especially the third state obtains a significant enhancement relative to the initial height, and also a reduction in energy. The energy shifts become smaller for the higher states, reflecting that the probability of excitations and the influence of the interaction is minimal at these energy scales. 
The effect of increasing $\eta$ is illustrated in the lower panel of Fig.~\ref{fig:pSpg_f-05et1}, where we have shown results after 15 iterations of calculations with $\eta=0.5,1$ and 7 in the $N=10$, $p=4$ system with $f=-0.5$. The $\eta=7$ results have a closer resemblance to a calculation with weaker interaction strength (lower absolute value of $|f|$). The energy levels are more evenly spaced and more of the strength is conserved in the lowest levels, illustrating the general effect of $\eta$ as a parameter which lessens the effect of the interaction and forces the solution closer to a mean-field pattern. 

\subsection{Comparison with exact diagonalization}\label{subsec:compres}

\begin{figure}[hbtp]
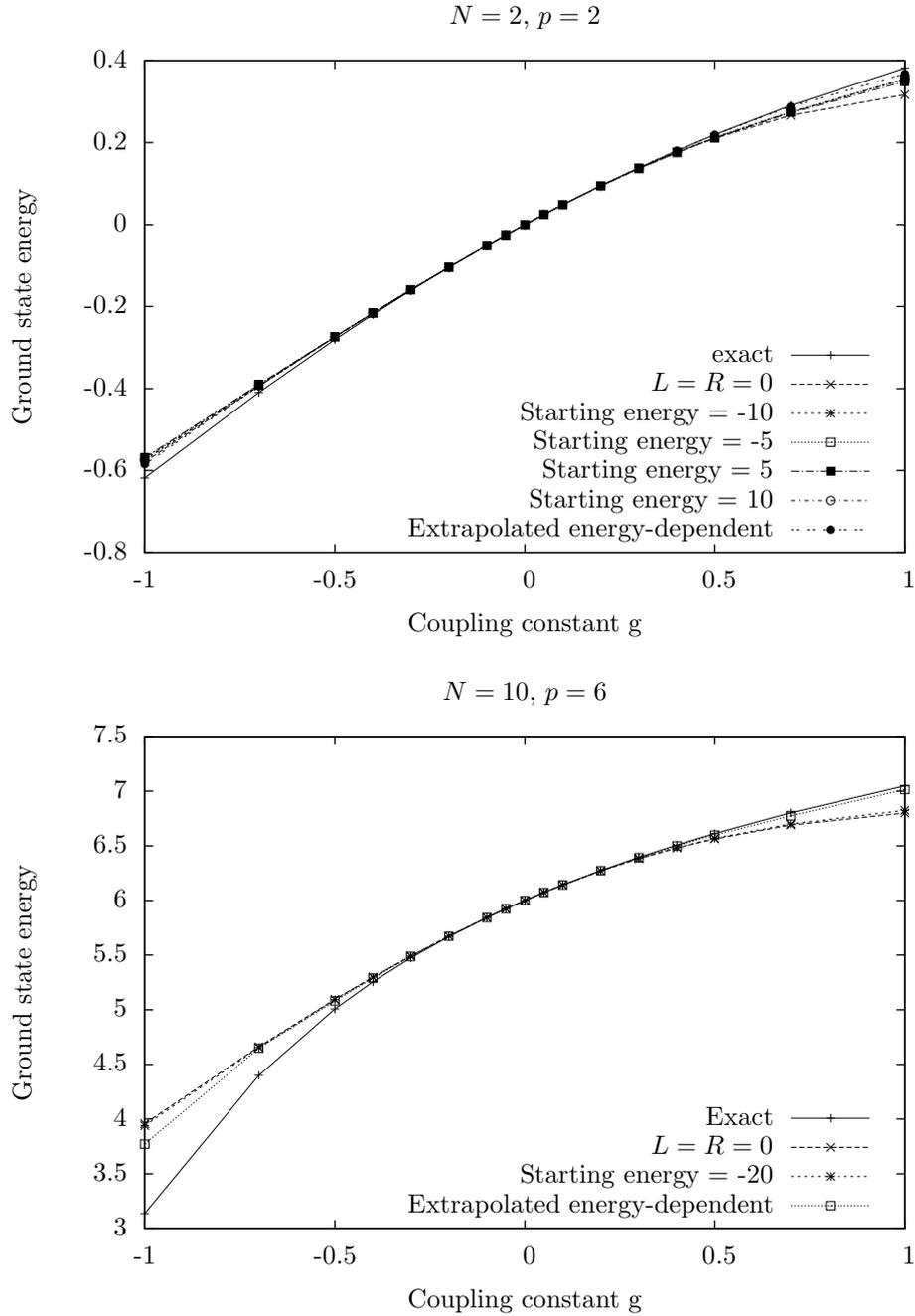

\begin{tabular}{c}
\input{gn2p2.tex} \\ 
\input{gn10p6.tex} \\ 
\end{tabular}
\caption{Comparison between the exact solution and the different Parquet solutions in the $N=2$, $p=2$ model (upper panel) and in the $N=10$, $p=2$ model (lower panel). The differences between the exact solution and the various Parquet solutions are much larger in the $N=10$ case. The energy-dependent results are slightly closer to the exact solution than the energy-independent results.} 
\label{fig:pg}
\end{figure}

We have compared the different approximation schemes discussed in Section~\ref{sec:itersolution} with the exact ground state energies obtained from diagonalization. We compare the ground state energies as a function of the ratio of interaction strength $g$ to the level spacing as the number of levels increase and as more particles are added. 

For convergent starting energies in an energy-independent calculation, the different starting energies yield minimal differences. We have chosen to work mostly with a starting energy of -20, as this is convergent for all values of $g$ we have studied. All convergent results for other starting energies have less than 3\% deviation from the results shown.

In the upper panel of Fig.~\ref{fig:pg} we show results for the simplest case, with  $N=2$ and $p=2$, as a function of coupling strength. We compare the ground state energy found from an exact solution, a calculation where $L=R=0$, an energy-independent calculation at different starting energies and extrapolated energy-dependent values. In the $L=R=0$ calculations we have performed a self-consistent calculation of the single-particle energies and binding energy based on the input interaction $V$ alone.

We see that the agreement between our results and the exact is very good for small values of the coupling strength $|g|$, where the independent-particle properties are dominant. For stronger negative coupling relative to the level spacing, the Parquet calculations underbind, probably due to the error introduced in our approximate solution of the Dyson equation. The differences between full Parquet and the results for $L=R=0$ are very small, indicating that contributions to the ground state energy are provided by the first- and second-order self energy diagrams in the Parquet solution of this simple model. The results for the starting energies $E_{in}= -10,-5,5$ and 10 demonstrate that in the pair-conserving model the dependence on $E_{in}$ is very small. 

 The energy-dependent results agree very well with the energy-independent data in the range of $g$ values where the calculations converge. To improve the convergent range, $\eta$ has to have a non-zero value, giving the graph in the upper panel of Fig.~\ref{fig:pg}. A calculation based on several values  of $\eta$ and linearly interpolated to $\eta=0$ performs slightly better at large $g$ values. The tendency to underbind at more negative $g$ values persists, showing that the energy dependence alone is probably not enough to recover all of the missing binding.

When we increase the number of levels to 10 and the number of particles to six, we obtain the graphs in the lower panel of Fig.~\ref{fig:pg}. We see increasing differences between the exact values and the Parquet calculated values at stronger negative coupling. The extrapolated values are still better than the energy-independent values which in turn are slightly better than the $L=R=0$ results. The dependence on $E_{in}$ is still negligible, which is why only the $E_{in}=-20$ graph is shown. 

\begin{figure}[hbtp]
\input{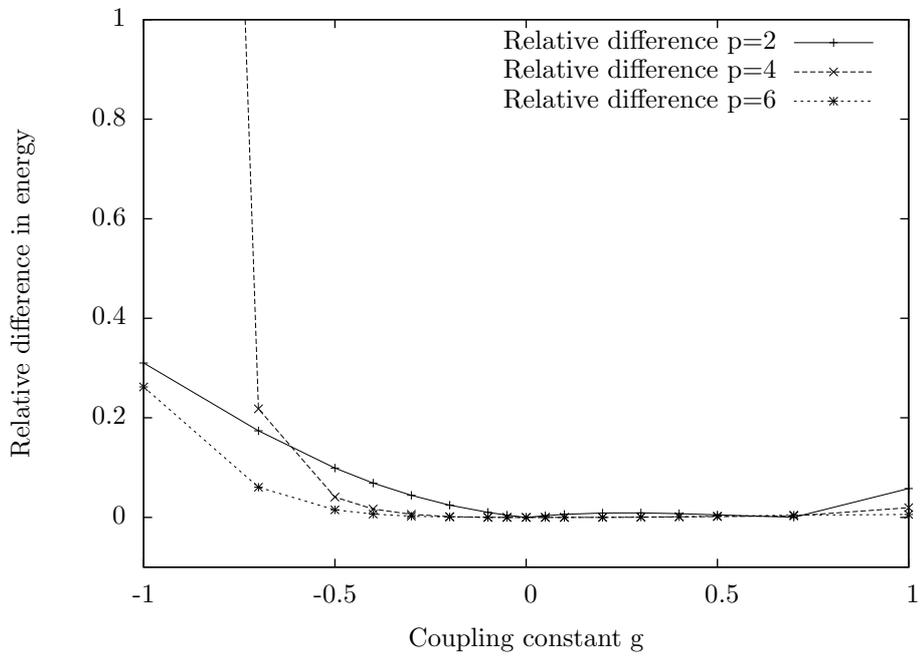}
\caption{The relative difference in energy $\varepsilon = |E^{\mathrm{Parquet}}-E^{\mathrm{Exact}}|/|E^{\mathrm{Exact}}|$ for $p=2$, $p=4$ and $p=6$.} 
\label{fig:pg_n10ad}
\end{figure}
To study the effects of increasing particle number, we show in Fig.~\ref{fig:pg_n10ad} the absolute difference 
\begin{equation*}
\varepsilon = \frac{|E^{\mathrm{Parquet}}-E^{\mathrm{Exact}}|}{|E^{\mathrm{Exact}}|}
\end{equation*} 
between our energy-dependent results and the results from the exact diagonalization for the $N=10$ model with two, four and six particles. The odd data point at $f=-1$ in the $p=4$ graph has a value close to 7. This rapid increase is mainly due to the fact that the exact value at this point is close to 0, amplifying the difference at this point. We see that in general, the relative errors of our results reduce as the number of particles increase. 

\begin{figure}[hbtp]
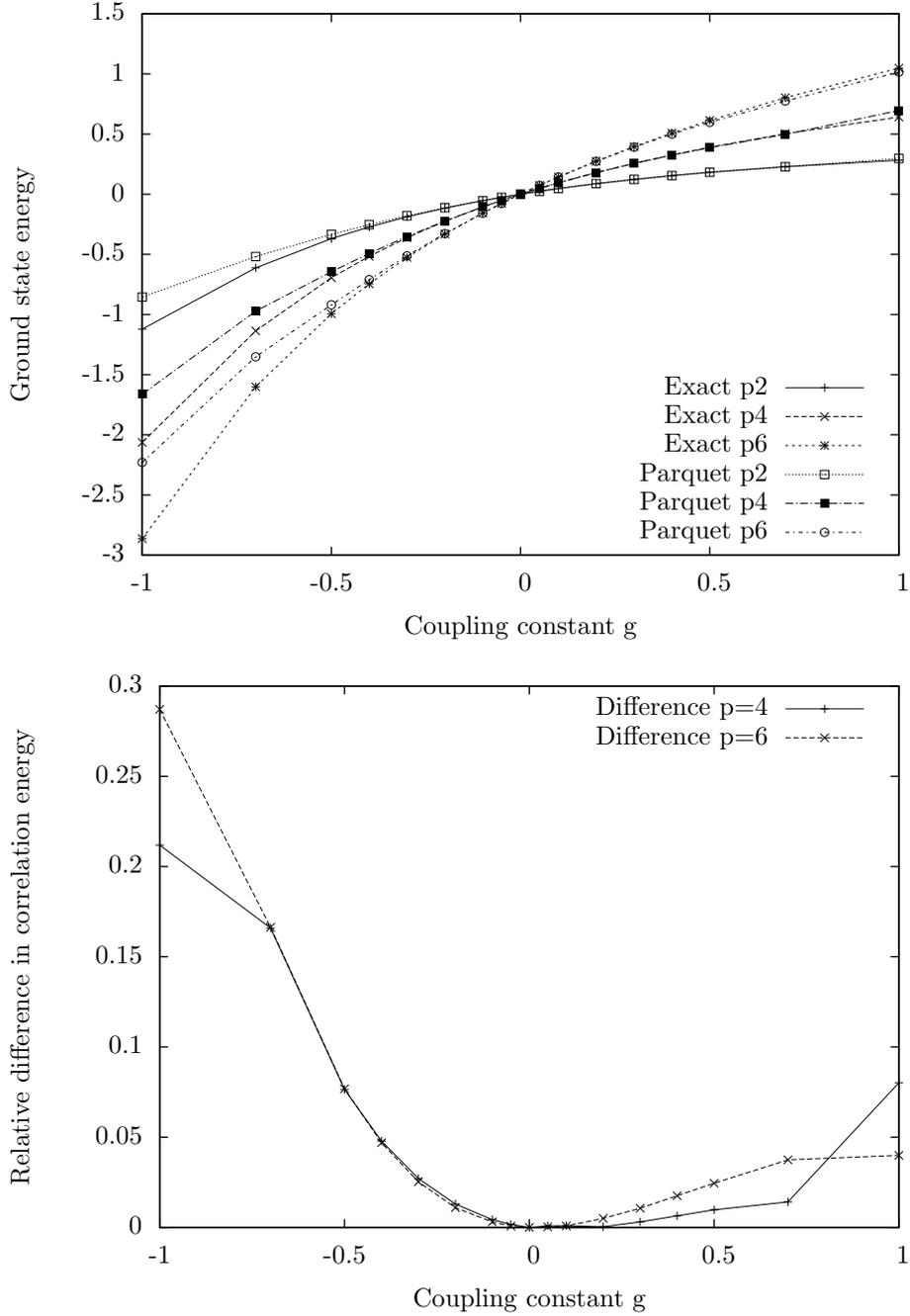

\begin{tabular}{c}
\input{gn10.tex}\\
\input{gn10diff.tex}\\
\end{tabular}
\caption{Comparison between the $N=10$, $p=2$ and adjusted $p=4$, $p=6$ exact and Parquet energy-dependent results are shown in the upper panel. The lower panel shows the relative difference in correlation energy $\varepsilon_C = |E_C^{\mathrm{Parquet}}-E_C^{\mathrm{Exact}}|/|E^{\mathrm{Exact}}_C|$ for $p=4$ and $p=6$.} 
\label{fig:pg_n10comp}
\end{figure}

We have also investigated the changes in correlation energy as the number of particles change.  We show in Fig.~\ref{fig:pg_n10comp} a comparison between the energy-dependent results and exact diagonalization results. The two latter graphs are adjusted such that the ground state energies at $g=0$ are equal. Thus the graphs for $p=4$ and $p=6$ show the correlation energy $E_C$ in the system due to the increased number of particles. As the number of particles increase, the correlation energy changes more rapidly with increasing interaction strength, as could be expected.

In the lower panel we show the relative difference in correlation energy
\begin{equation*}
\varepsilon_C = \frac{|E_C^{\mathrm{Parquet}}-E_C^{\mathrm{Exact}}|}{|E^{\mathrm{Exact}}_C|}
\end{equation*}
for the $p=4$ and $p=6$ cases relative to the $p=2$. The errors have two distinct sources, namely the systematic errors introduced by the approximations employed in the solution, and the errors stemming from missing many-body correlations. If the error $|E_C^{\mathrm{Parquet}}-E_C^{\mathrm{Exact}}|$ scales as the exact correlation energy with respect to the number of particles, the relative difference would be independent of particle number for four and six particles. In the lower panel of Fig.~\ref{fig:pg_n10comp} we have plotted the relative difference. The graphs for four and six particles coincide in the range $-0.7<f<0$. The $f>0$ values show that the Parquet solution is very close to the exact, and the different behaviour might originate from the uncertainties in the extrapolation to $\eta{\rightarrow}0$.  

\begin{figure}[hbtp]
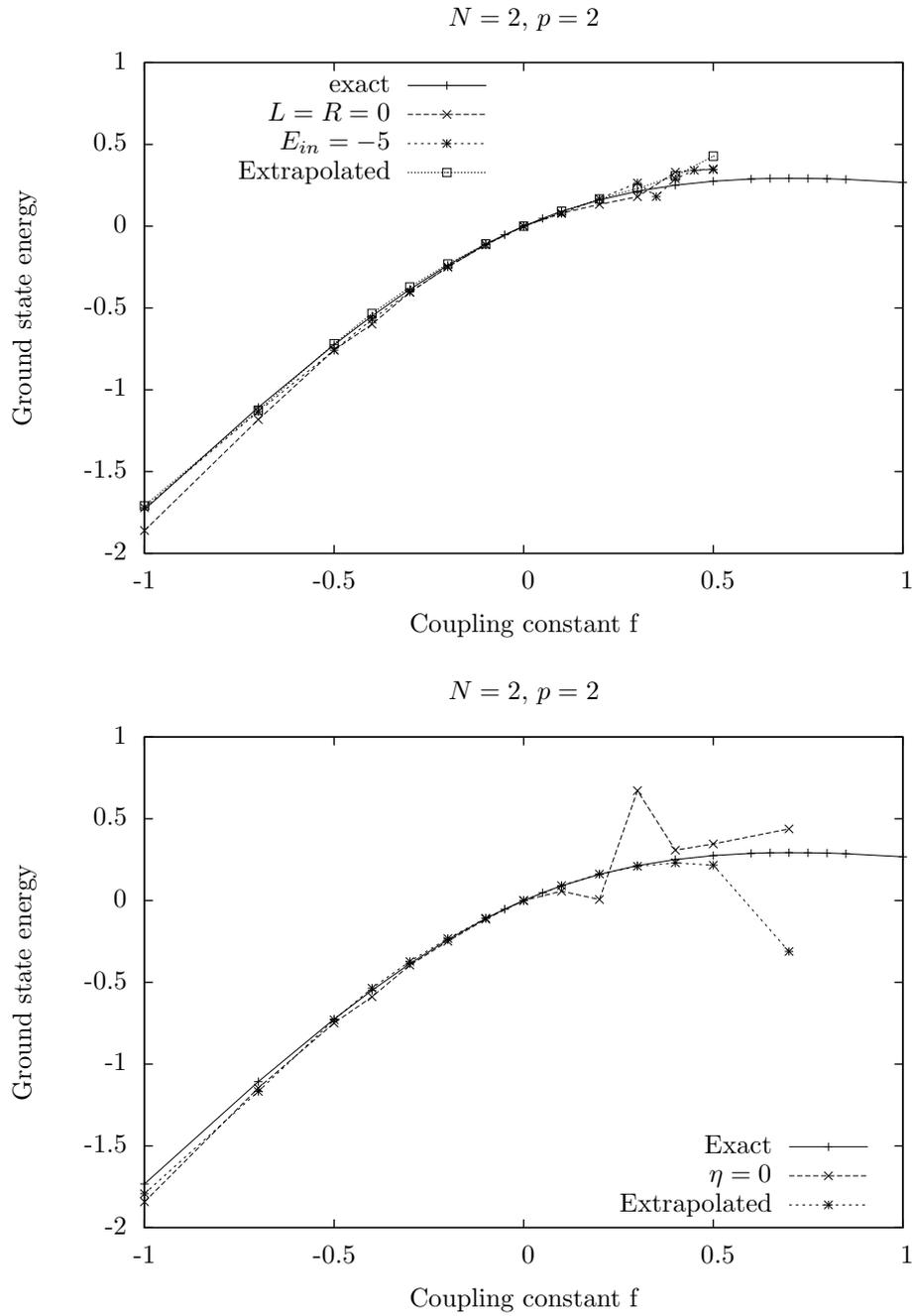

\begin{tabular}{c}
\input{fhfn2p2_ed1.tex} \\ 
\input{fhfn2p2_ed20.tex} \\ 
\end{tabular}
\caption{Ground state energy of calculations in the pair-breaking model compared with exact solution for the $N=2$, $p=2$ system. The results in the upper panel are for energy-independent calculations, while the lower panel shows energy-dependent results. For negative values of $f$, the agreement between our calculations and the exact solution is very good. Our calculations become increasingly unstable for positive values of $f$.} 
\label{fig:fhf_n2p2}
\end{figure}

In the pair-breaking model, calculations for the simple $N=2$, $p=2$ system have a close correspondence to the exact calculation for $f<0$ values, as shown in Fig.~\ref{fig:fhf_n2p2}. The differences between different starting energies are rather small in this model as well. The upper panel of the figure shows graphs for $E_{in}=-5$ and an extrapolation based on extrapolation from calculations with $\eta=1,5$ and 10. For $f>0$ the results destabilize around $f=0.3$, as expected from the discussion on the convergence. The extrapolated results are not able to do any better than the $\eta=0$ results, and overbinds at $f<0$ in the same manner as the $L=R=0$ results. Energy-dependent calculations, as shown in the lower panel of Fig.~\ref{fig:fhf_n2p2}, show in general the same behaviour as the energy-independent calculations. Results extrapolated from calculations with $\eta=1,2,3,5$ and 10 manage to agree nicely with the exact solution up to $f=0.5$ before becoming too unstable.
 \begin{figure}[hbtp]
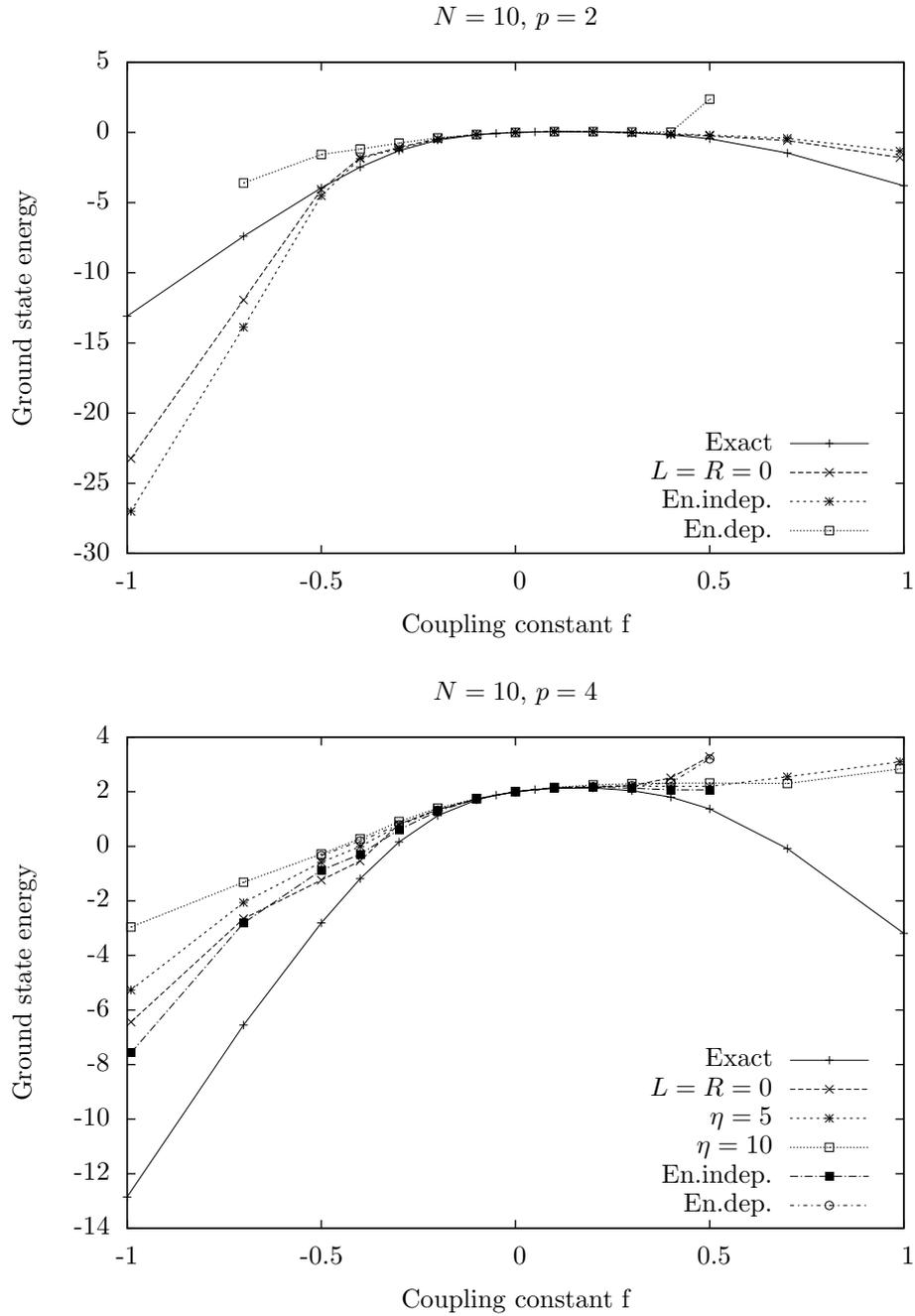

\begin{tabular}{c}
\input{fhfn10p2.tex} \\ 
\input{fhfn10p4.tex} \\ 
\end{tabular}
\caption{Ground state energy of calculations compared with exact solution for the $N=10$, $p=2$ model (upper panel) and $N=10$, $p=4$ (lower panel). The results shown are convergent, but unconverged results for lower $\eta$ values indicate that the extrapolations probably do not reflect the correct $\eta{\rightarrow}0$ limit.} 
\label{fig:fhf_n10}
\end{figure}
As stated in Section \ref{subsec:pair_f_eta}, all calculations become unstable at $f=1$. It is possible to find convergent solutions for values of $f>1$, but they are very far from the exact value. The exact ground state energy decreases, while the Parquet solutions increase. The failure is due to the reversal of the lowest-lying levels, violating our assumption that the input basis should not differ too radically from the expected solution. 

The results for $N=10$, $p=2$ and $N=10$, $p=4$ are shown in Fig.~\ref{fig:fhf_n10}. We compare $L=R=0$, energy-independent and energy-dependent calculation for calculations with the exact solution for both the $N=10$, $p=2$ model (upper panel) and $N=10$, $p=4$ (lower panel). 

In the $N=10$, $p=2$ model we see that the instability around $f=-0.5$ seems to indicate a breakdown, as both the $L=R=0$ and the energy-independent calculations give far too negative ground state energies when $f<-0.5$. In the $N=10$, $p=4$ model the energy-independent calculations and the energy-dependent calculations are rather close, and slightly closer to the exact solution than the $L=R=0$ calculations. 

All values shown in Fig.~\ref{fig:fhf_n10} are converged, but the unconverged results at smaller $\eta$ values show oscillations between a state with larger and a state with a much lower ground state energy than the exact value. The $N=10$, $p=2$ results shown are consistent with the lowest energy, while the $N=10$, $p=4$ results is more in line with the highest energy in these (large) fluctuations. This might indicate that the extrapolated results shown does not represent the ground state energy at $\eta=0$ properly in either system. The discrepancies become significant at values $f\lesssim{-0.5}$, consistent with the indications of a breakdown of our solutions around this interaction strength. 

It must be remembered that all values are extrapolated and carry an uncertainty which roughly estimated is larger than the observed differences between the three approaches. Thus it is more prudent to look at the general trends rather than analyze the finer points in any great detail.

\section{Conclusion} \label{sec:conclusions}

We have applied our implementation of the Parquet method to a simplified model of the nucleus, a model with a constant spacing between single-particle levels and particles interacting via a constant interaction. We have looked at two schemes of the interaction, one in which only pair-excitations are allowed, and one with a pair-breaking term the same size as the pair-conserving term. In both cases we have investigated the stability with respect to $\eta$ when varying different parameters, and compared different energy approximation schemes to the exact solution found by matrix diagonalization. 

We have found that the Parquet method performs generally well with the pair-conserving interaction. In this model the self energy is diagonal. The stability of our solutions depend on the parameter $\eta$, which regulates the influence of the pole structures in the propagators. The convergence properties with respect to $\eta$ are normally quite good, with some notable exceptions when a pole in the generated interaction $\Gamma$ is encountered. At small system sizes, the number of poles is sufficiently small so that no $\eta$ is needed, but as the number of levels increases, a finite, but small $\eta$ is needed. The number of particles have little impact on the convergence properties.

The energy-dependent scheme was found to perform slightly better than the energy-independent scheme when compared with the exact results. Over a range of interaction strengths between -0.5 and 0.5 of the level spacing both schemes show good agreement with the exact solution, also as the number of levels and particles in the model is increased. At larger interaction strengths, our solutions underbind the systems. In the pair-conserving model, the effect of increasing the number of available levels is rather large. It is difficult to ascertain the relative importance of the systematic errors stemming from the limitations imposed by our approximations when implementing the Parquet method versus the effect of missing many-body correlations. The study of the correlation energy with increasing particle number indicate that our solution method scale well with increasing particle number.

Introducing a pair-breaking force destabilizes the solutions in the region of strong repulsion ($f\gtrsim{0.5}$) in the smallest system with two levels and two particles, with a complete breakdown of the method when the first-order energies of the lowest levels become equal. Increasing the number of levels in the model gives larger differences between the exact solution and the Parquet solutions as the absolute value of the interaction strength increase. Increasing the number of particles gives increasingly larger discrepancies. 

As discussed, the energy-dependent calculation method is closer to the exact solution in the pair-conserving systems, but this is not the case in the pair-breaking systems. To obtain convergence in these latter systems, $\eta$ has to be increased to such large values that the solution becomes a mean-field-type solution with an almost energy-independent self energy. The differences between results for different $\eta$ values are small. Most of the correlations have been lost, and the energy-dependent results are rather far away from the exact solution. 

The energy-independent calculation method gives results closer to the exact solution than the energy-dependent scheme. Within this simple model, there is almost no dependence on the starting energy $E_{in}$, as long as this is chosen at values where the generated interaction $\Gamma$ does not have any poles.

The general conclusion is that for the larger pair-breaking systems, the Parquet method as implemented is only in agreement with the exact solution for $-0.3\lesssim{f}\lesssim{0.2}$. 

The main conclusion of our study of the Parquet method applied to our simple model is that its best area of application  is when the interaction strength is rather small compared to the level spacing. The method seems to fail to find solutions close to the exact solution as the number of levels and particles increase. The energy-dependent scheme is closer to the exact solution for small systems where the number of poles are limited, but as the number of poles in both self energy and generated interaction $\Gamma$ increase when the system size increase, this solution scheme needs large values of $\eta$ to converge. Then too many details of the pole structure are lost, and the solutions found by this scheme are further removed from the exact solution than the energy-independent solution.

In a larger perspective, the Parquet method can be implemented and give reasonable results which seems on the right track to becoming useful as an {\it ab initio} nuclear structure calculation method. The simple model discussed above has many factors in common with realistic systems. Many of the general conclusions and insights gained can be transferred to an application to real nuclei. It is easy to generate graphs for the self energy as a function of energy, and these can be used to determine the degree to which the solution incorporates correlations beyond a mean-field type solution. The spectral functions are easy to extract, as are spectroscopic factors.

The Parquet method certainly has the possibility for attaining a high level of accuracy. The angular-momentum coupling schemes allow for huge reductions in the model space, giving the possibility to perform much larger calculations. All Goldstone ground state energy diagrams to fourth order are included, a feat which in the coupled-cluster approach require inclusion of excitation operators to the fourth order (implying a CCSDTQ calculation)~\cite{Bartlett1981}. The results on the ground state energy of ${}^4$He~\cite{Barbieri2009b} show that the required level of precision to meet the current standard is possible within implementations of Green's function based methods. Results for nuclei like 
${}^4$He and $^{16}$O will be presented in a forthcoming work \cite{elise2010}.



\begin{thebibliography}{200}


\bibitem{coester}
F.~Coester, Nucl.~Phys.~7 (1958) 421. 

\bibitem{coesterkummel}
F.~Coester and H.~K{\" u}mmel, Nucl.~Phys.~17 (1960) 477.


\bibitem{Bartlett2007}
R.J.~Bartlett and M.~Musia\l, Rev.~Mod.~Phys.~79 (2007) 1291.

\bibitem{bishop97}
R.F~. Bishop, in Microscopic Quantum Many-Body
Theories and Their Applications, edited by J. Navarro and A.
Polls, Lecture Notes in Physics, Vol. 510, Springer, Berlin, 1998, p. 1.

\bibitem{Dean2004a}
D.~J. Dean and M.~Hjorth-Jensen, Phys. Rev. C 69 (2004) 054320.

\bibitem{Hagen2007}
G.~Hagen, D.~J. Dean, M.~Hjorth-Jensen, T.~Papenbrock, and A.~Schwenk, Phys.~Rev.~C 76 (2007) 044305.

\bibitem{Hagen2008}
G.~Hagen, T.~Papenbrock, D.~J. Dean, and M.~Hjorth-Jensen, Phys.~Rev.~Lett.~101 (2008) 092502.

\bibitem{Varga1995}
K.~Varga and Y.~Suzuki, Phys.~Rev.~C 52 (1995) 2885.

\bibitem{Viviani2005}
M.~Viviani, A.~Kievsky, and S.~Rosati, Phys.~Rev.~C 71 (2005) 024006.

\bibitem{Caurier2005}
E.~Caurier, G.~Mart\'inez-Pinedo, F.~Nowacki, A.~Poves, and A.P.~Zuker, Rev. Mod. Phys.~77 (2005) 2427.

\bibitem{Gazit2006}
D.~Gazit, S.~Bacca, N.~Barnea, W.~Leidemann, and G.~Orlandini, Phys.~Rev.~Lett.~96 (2006) 112301.

\bibitem{Pudliner1997}
B.S.~Pudliner, V.R.~Pandharipande, J.~Carlson, S.C.~Pieper, and R.B.~Wiringa, Phys.~Rev.~C 56 (1997) 1720.

\bibitem{Koonin1997}
S.E.~Koonin, D.J.~Dean, and K.~Langanke, Phys.~Rep.~278 (1997) 1.

\bibitem{Ceperley1995}
D.M.~Ceperley, Rev.~Mod.~Phys.~67 (1995) 279.

\bibitem{Nogga2000}
A.~Nogga, H.~Kamada, and W.~Gl\"ockle, Phys.~Rev.~Lett.~85 (2000) 944.

\bibitem{Arriaga1995}
A.~Arriaga, V.R.~Pandharipande, and R.B.~Wiringa, Phys.~Rev.~C 52 (1995) 2362.

\bibitem{Barbieri2006}
C.~Barbieri, Phys.~Lett.~B643 (2006) 268.

\bibitem{Barbieri2009}
C.~Barbieri and W.H.~Dickhoff, Int.~J.~Mod.~Phys.~A 24 (2009) 2060.

\bibitem{Lapikas1993}
L.~Lapik{\'a}s, Nucl.~Phys.~A553 (1993) 297.

\bibitem{Leuschner1994}
M.~Leuschner, J.R.~Calarco, F.W.~Hersman, E.~Jans, G.J.~Kramer,
  L.~Lapik\'as, G.~van~der Steenhoven, P.K.A.~de~Witt~Huberts, H.P.~Blok,
  N.~Kalantar-Nayestanaki, and J.~Friedrich, Phys. Rev. C 49 (1994) 2955.

\bibitem{Subedi2008}
R.~Subedi {\em et al}, Science 320 (2008) 1476.

\bibitem{Rohe2004}
D.~Rohe {\em et al}, Phys.~Rev.~Lett.~93 (2004) 182501.

\bibitem{Barbieri2004a}
C.~Barbieri and L.~Lapik\'as, Phys.~Rev.~C 70 (2004) 054612.

\bibitem{Baym1961}
Gordon Baym and Leo~P. Kadanoff, Phys.~Rev.~124 (1961) 287.

\bibitem{Baym1962}
G.~Baym, Phys.~Rev.~127 (1962) 1391.

\bibitem{Diatlov1957}
I.T.~Diatlov, V.V.~Sudakhov, and K.A.~Ter-Martirosian, JETP 5 (1957) 631.

\bibitem{Zheleznyak1997}
A.T.~Zheleznyak, V.M.~Yakovenko, and I.E.~Dzyaloshinskii, Phys.~Rev.~B 55(1997) 3200.

\bibitem{Janis2006}
V.~Janis, Cond.~Matt.~Phys.~9 (2006) 499.

\bibitem{Bickers1992}
N.E.~Bickers and D.J.~Scalapino, Phys.~Rev.~B 46(1992) 8050.

\bibitem{Yeo1996}
J.~Yeo and M.A.~Moore, Phys.~Rev.~B 54 (1996) 4218.

\bibitem{Jackson1982}
A.D.~Jackson, A.~Lande, and R.A.~Smith, Phys.~Rep.~86 (1982) 55.

\bibitem{Jackson1994}
A.D.~Jackson and T.~Wettig, Phys.~Rep.~237 (1994) 325.

\bibitem{Jackson1985}
A.D.~Jackson, A.~Lande, R.W.~Guitink, and R.A.~Smith, Phys.~Rev.~B 31 (1985) 1403.

\bibitem{Lande1992}
A.~Lande and R.A.~Smith, Physical Review A 45(1992) 2913.

\bibitem{Yasuda1999}
K.~Yasuda, { Physical Review A} 59 (199) 4133.

\bibitem{Ellis1977}
P.J.~Ellis and E.~Osnes, Rev.~Mod.~Phys.~49 (1977) 4777.

\bibitem{mhj1995}
M.~Hjorth-Jensen, T.T.S.~Kuo and E.~Osnes, Phys.~Rep.~261 (1995)  121. 

\bibitem{FetterWalecka1971}
A.L.~Fetter and J.D.~Walecka, Quantum Theory of Many-particle Systems, McGraw-Hill, New York, 1971.

\bibitem{Mattuck1976}
R.D.~Mattuck, A Guide to Feynman Diagrams in the Many-Body Problem, Dover Publications, 1976.

\bibitem{BlaizotRipka1986}
J.-P.~Blaizot and G.~Ripka, Quantum Theory of Finite Systems, MIT Press, Cambridge, MA, 1986.

\bibitem{Abrikosov1965}
A.A.~Abrikosov, L.P.~Gorkov, and I.E.~Dzjaloshinski, Quantum Field Theoretical Methods in Statistical Physics, Pergamon Press, Oxford, 1965.

\bibitem{Dickhoff2005}
W.H.~Dickhoff and D.~Van Neck, Many-body Theory exposed!: Propagator Description of Quantum
  Mechanics in Many-body Systems, World Scientific, Hackensack, NJ, 2005.

\bibitem{Dickhoff2004}
W.H.~Dickhoff and C.~Barbieri, Prog.~Part.~Nucl.~Phys.~52 (2004) 377.

\bibitem{Martin1959}
P.C.~Martin and J.~Schwinger,  Phys. Rev.~115 (1959) 1342.

\bibitem{boffi1971}
S.~Boffi, Lett.~Nuovo Cimento 1 (1971) 931.

\bibitem{Kuo1981}
T.T.S.~Kuo, J.~Shurpin, K.C.~Tam, E.~Osnes, and P.J.~Ellis, Ann.~of Phys.~132 (1981) 237.

\bibitem{Sakurai1994}
J.J.~Sakurai, Modern Quantum Mechanics, Addison-Wesley, Reading, MA, 1994.


\bibitem{Bartlett1981}
R.J~Bartlett, Ann.~Rev.~Phys.~Chem.~32 (1981) 359.

\bibitem{Barbieri2002a}
C.~Barbieri, Self-consistent Green's function study of low-energy
  correlations in $^{16}$O, PhD thesis, Washington University, 2002.

\bibitem{Barbieri2001}
C.~Barbieri and W.H. Dickhoff, Phys.~Rev.~C 63 (2001) 34313.

\bibitem{Barbieri2002}
C.~Barbieri and W.H.~Dickhoff, Phys.~Rev.~C 65 (2002) 064313.

\bibitem{Barbieri2003}
C.~Barbieri and W.H.~Dickhoff, Phys.~Rev.~C 68 (2003) 014311.

\bibitem{Barbieri2004}
C.~Barbieri, C.~Giusti, F.D.~Pacati, and W.H.~Dickhoff, Phys.~Rev.~C, 70 (2004) 014606.

\bibitem{Talmi1993}
I.~Talmi, Simple Models of Complex Nuclei: the Shell Model and
  Interacting Boson Model, Harwood Academic Publishers, 1993.

\bibitem{Dean2003}
D.J.~Dean and M.~Hjorth-Jensen, Rev.~Mod.~Phys.~75 (2003) 607.


\bibitem{Barbieri2009b}
C.~Barbieri, Phys.~Rev.~Lett.~103 (2009) 202502

\bibitem{elise2010}
E.~Bergli and M.~Hjorth-Jensen, in preparation (2010).
\end{thebibliography}
\end{document}